\documentclass[aps,prb,reprint,twocolumn,superscriptaddress,showpacs]{revtex4-1}
\usepackage{amsmath}
\usepackage{amssymb}
\usepackage{bbold}
\usepackage{bm}
\usepackage{braket}
\usepackage{graphicx}
\usepackage{hyperref}

\newcommand{\up}{\uparrow}
\newcommand{\down}{\downarrow}
\newcommand{\loc}{\mathrm{loc}}
\newcommand{\nonloc}{\mathrm{nl}}

\begin{document}
    \title{Signatures of nonlocal Cooper-pair transport and of a 
        singlet--triplet transition 
        \\ in the critical current of a double-quantum-dot Josephson junction}

    \author{B.~Probst}
    \affiliation{Institut f\"ur Mathematische Physik, Technische Universit\"at
        Braunschweig, D-38106 Braunschweig, Germany}
    \author{F.~Dom\'inguez}
    \affiliation{Departamento de F\'isica Te\'orica de la Materia Condensada,
        Condensed Matter Physics Center (IFIMAC), \\ and Instituto Nicol\'as
        Cabrera, Universidad Aut\'onoma de Madrid, E-28049 Madrid, Spain}
    \affiliation{\mbox{Institut f\"ur Theoretische Physik und Astrophysik, 
        Universit\"at W\"urzburg, D-97074  W\"urzburg, Germany}}
    \author{A.~Schroer}
    \affiliation{Institut f\"ur Mathematische Physik, Technische Universit\"at
        Braunschweig, D-38106 Braunschweig, Germany}
    \author{A.~Levy Yeyati}
    \affiliation{Departamento de F\'isica Te\'orica de la Materia Condensada,
        Condensed Matter Physics Center (IFIMAC), \\ and Instituto Nicol\'as
        Cabrera, Universidad Aut\'onoma de Madrid, E-28049 Madrid, Spain}
    \author{P.~Recher}
    \affiliation{Institut f\"ur Mathematische Physik, Technische Universit\"at
        Braunschweig, D-38106 Braunschweig, Germany}
    \affiliation{Laboratory for Emerging Nanometrology Braunschweig, D-38106 
        Braunschweig, Germany}

    \begin{abstract}
        We study the critical Josephson current flowing through a double quantum
        dot weakly coupled to two superconducting leads. We use analytical as
        well as numerical methods to investigate this setup in the limit of
        small and large bandwidth leads in all possible charging states, where
        we account for on-site interactions exactly. Our results provide clear
        signatures of nonlocal spin-entangled pairs, which support
        interpretations of recent experiments [Deacon, R. S.  et al., Nat.\
        Commun.\ 6, 7446 (2015)]. In addition, we find that the ground state
        with one electron on each quantum dot can undergo a tunable
        singlet--triplet phase transition in the regime where the
        superconducting gap in the leads is not too large, which gives rise to
        an additional new signature of nonlocal Cooper pair transport.
    \end{abstract}

    \date{\today}

    % 74.50.+r 	Tunneling phenomena; Josephson effects (for SQUIDs, see 
    %           85.25.Dq; for Josephson devices, see 85.25.Cp; for Josephson 
    %           junction arrays, see 74.81.Fa)
    %74.45.+c 	Proximity effects; Andreev reflection; SN and SNS junctions
    %03.65.Ud 	Entanglement and quantum nonlocality (e.g. EPR paradox, Bell's 
    %           inequalities, GHZ states, etc.) (for entanglement production and
    %           manipulation, see 03.67.Bg; for entanglement measures, witnesses
    %           etc., see 03.67.Mn; for entanglement in Bose-Einstein 
    %           condensates, see 03.75.Gg)

    \maketitle

    \section{Introduction}
        \begin{figure}
            \includegraphics{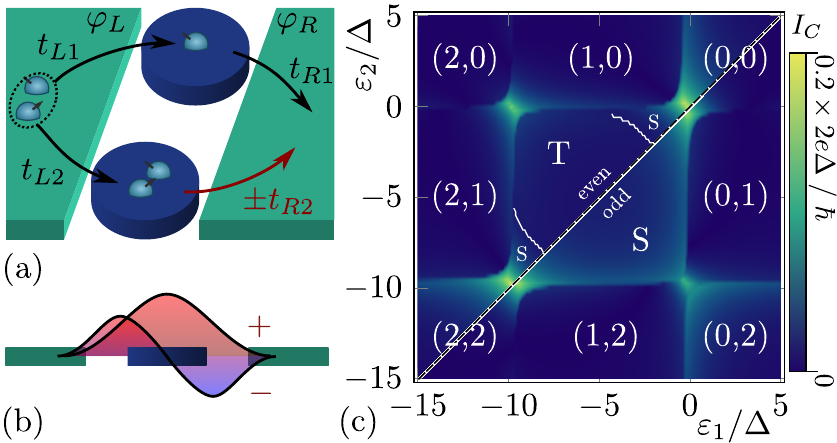}
            \caption{(Color online) Double-quantum-dot Josephson junction. 
                (a) The Josephson current is carried by Cooper pairs which 
                tunnel coherently between two superconducting leads with 
                superconducting phases $\varphi_L$ and $\varphi_R$.
                Microscopically, this involves four single-particle tunneling
                events with amplitudes $t_{\nu i}$. Local transport (both
                electrons of a Cooper pair tunnel through a single quantum dot)
                can be distinguished from nonlocal transport (the two electrons
                of a Cooper pair tunnel through different quantum dots).
                (b) The symmetry of the orbital wave functions on the quantum
                dots is captured in the total tunnel parity 
                $\mathcal{P}=\text{sgn}(t_{L1}t_{L2}t_{R1}t_{R2})$, or,
                equivalently, in $\pm t_{R2}$, and has distinctive signatures in
                the critical current.
                (c) Critical current across the junction at zero temperature as 
                a function of the quantum-dot level energies $\varepsilon_{1,2}$
                obtained in the zero-bandwidth approximation
                (Sec.~\ref{sec-zbw}).  The upper-left half of the plot shows the
                critical current at \emph{even} tunnel parity and the
                lower-right half at \emph{odd} tunnel parity. The critical
                current becomes large close to ground-state transitions where
                the charge of the quantum dots $(N_1,N_2)$ fluctuates. At even
                tunnel parity a transition between a nonlocal singlet (S) and a
                triplet (T) ground state in the $(1,1)$ sector emerges. The
                parameters are $|t|=0.5\Delta$ and the Coulomb repulsion
                $U=10\Delta$, where $\Delta$ is the magnitude of the
                superconducting gap in the leads.}
            \label{fig-setup}
        \end{figure}

        Creating mobile spin-entangled electron pairs in solid state transport
        setups has been the subject of intensive research in recent years.
        \cite{Recher01, Lesovik01, Recher02, Bena02, Recher03, Prada04,
        Yamaguchi02, Yeyati07, Cayssol08, Sato10, Schroer15b} The proposed
        setup of Ref.~\citenum{Recher01} consists of an $s$-wave superconductor
        coupled to two quantum dots (QDs) in the Coulomb-blockade regime, which
        are further coupled to outgoing Fermi liquid leads, \cite{Hofstetter09,
        Herrmann10, Das12, Tan15, Fulop15} where a dominant transport channel
        comprising pairwise and nonlocal transport has been identified.  The
        degree of spin-entanglement of the nonlocal pairs, however, has so far
        not been measured as it is not straight forward to measure spin
        correlations directly. Many ways have been proposed to detect the spin
        entanglement ranging from violating a Bell inequality, \cite{Kawabata01,
        Chtchelkatchev02, Braunecker13} noise properties in a beam-splitter
        setup \cite{Burkard00, Hu04, Samuelsson04-2, Prada06, Mazza13,
        Schroer14} or converting the electron-spin pairs to photons.
        \cite{Cerletti05, Titov05, Budich09, Nigg14, Schroer15} Recently,
        Deacon et al.~\cite{Deacon15} realized a Josephson junction containing a
        double quantum dot (DQD) embedded in the junction.  The critical current
        of the junction was investigated and signs of nonlocal pair transport
        coherently involving both QDs were conjectured by showing that the
        results are incompatible with two uncorrelated transport channels of
        supercurrent separately via each QD.

        Here, we analyze in detail the setup of Ref.~\citenum{Deacon15} using
        both analytical and numerical tools. We thereby go beyond existing
        theoretical work on the proposed setup \cite{Choi00, Wang11} by taking
        into account all of the filling factors of the QDs up to two 
        levels per QD. We also relax the assumption of a large superconducting
        gap compared to the QD level energies $\varepsilon_i$ and
        the charging energy $U_i$ of the QDs $i=1,2$. The signatures of nonlocal
        pair transport we investigate do not rely on the Aharanov--Bohm effect
        \cite{Pan06,Wang11,Jacquet15} or on SQUIDs. \cite{Choi00,Wang11}

        We first introduce the model Hamiltonian in Sec.~\ref{sec-model} and a
        special limit, the zero-bandwidth approximation, where the
        superconducting leads are represented by a single site with pairing
        interaction.  We then calculate the critical current in the ground state
        of the junction numerically by exact diagonalization and compare it to a
        fourth-order perturbation theory in the tunneling from the DQD to the
        superconducting leads, both in the zero-bandwidth limit and in the
        wide-band limit. In the perturbation theory, the different tunneling
        paths, local and nonlocal contributions, become explicit. The ground
        state in the $(1,1)$-charging sector of the DQD (Figs.~\ref{fig-setup}
        and~\ref{fig-ic}) changes as a function of the level energies on the QDs
        and/or charging energy from nonlocal singlet to nonlocal triplet. The
        triplet ground state is stabilized in the regime
        $\varepsilon_i/|\Delta_\nu|> 1$ by cotunneling processes, which are
        independent of the superconducting phase difference and therefore are
        not of the Andreev reflection type. We show that the so-far undiscovered
        triplet phase also crucially depends on the parity of the tunneling
        matrix elements (Figs.~\ref{fig-setup} and~\ref{fig-ic}). Going beyond
        the zero-bandwidth approximation by considering superconducting leads in
        the wide-band limit, we show perturbatively in the tunneling, that the
        triplet phase also exists in this case if the charging energies $U_i$ of
        the QDs are finite. This extends the model considered in
        Refs.~\citenum{Choi00} and~\citenum{Wang11}, where only singlet phases
        in the ground state are predicted, to the case of finite $U_i$ hosting
        both singlet and triplet ground states.

        We further study a multilevel model numerically in which one of the QDs 
        has an additional single-particle level, and show that this 
        configuration gives a qualitatively good description of the experimental
        results of Ref.~\citenum{Deacon15}, assuming different parities for the
        tunneling matrix elements for the two levels on one QD.

        Finally, we present the analytical case of large superconducting gaps,
        where we can integrate out the superconducting leads obtaining an
        effective model for the DQD system only. This parameter regime
        corresponds to the limit where the nonlocal processes can be maximized
        by having, in addition, a charging energy much larger than the induced
        superconducting gap. Here, as expected, we obtain only a singlet ground
        state in the $(1,1)$ sector.

    \section{\label{sec-model}Model}
        We consider the geometry depicted in Fig.~\ref{fig-setup}(a). Two
        QDs $i=1,2$ are tunnel coupled in parallel to two $s$-wave
        superconductors $\nu=L,R$ at $\mathbf{x}=0$ with amplitudes $t_{\nu i}$,
        which are chosen real in the absence of a magnetic field.  Each QD
        contains only a single spin-degenerate level,
        $\sigma=\uparrow,\downarrow$, with energy $\varepsilon_i$ and with the
        local Coulomb repulsion $U_i$, which is relevant for transport.  There
        is no direct cross-talk between the QDs or between the superconductors.
        The Hamiltonian is
        \begin{equation}
            H=H_1+H_2+H_L+H_R+H_T
            \label{eq-h}
        \end{equation}
        with the QD contributions
        \begin{equation}
            H_i=\sum_\sigma\varepsilon_i d_{i\sigma}^\dagger d_{i\sigma}
                +U_i d_{i\uparrow}^\dagger d_{i\downarrow}^\dagger
                d_{i\downarrow} d_{i\uparrow},
        \end{equation}
        the superconducting lead contributions
        \begin{equation}
            H_{\nu}=\sum_{\mathbf{k}\sigma}\varepsilon_{\mathbf{k}\nu}
                c_{\mathbf{k}\nu\sigma}^\dagger 
                c_{\mathbf{k}\nu\sigma}
                +\sum_{\mathbf{k}}\Delta e^{-i\varphi_\nu} 
                c_{\mathbf{k}\nu\uparrow}^\dagger
                c_{-\mathbf{k}\nu\downarrow}^\dagger
                +\text{H.c.},
            \label{eq-hlead}
        \end{equation}
        and their tunnel coupling
        \begin{align}
            H_T&=\sum_{i\sigma\nu}t_{\nu i}
                d_{i\sigma}^\dagger\psi_{\nu\sigma}(0)+\text{H.c.} \notag\\
                &=\sum_{i\mathbf{k}\sigma\nu}t_{\nu i}
                d_{i\sigma}^\dagger c_{\nu\mathbf{k}\sigma}
                +\text{H.c.},
            \label{eq-ht}
        \end{align}
        where the $d_{i\sigma}$ operators annihilate electrons localized on the
        QDs and where the $c_{\mathbf{k}\nu\sigma}$ and the
        $\psi_{\nu\sigma}(\mathbf{x})$ operators annihilate spin-$\sigma$
        electrons in lead $\nu$ with momentum $\mathbf{k}$ or at position
        $\mathbf{x}$, respectively. The normal-state dispersion in the leads is
        $\varepsilon_{\mathbf{k}\nu}$. We assume the two superconducting leads
        to be of the same material with the same superconducting energy gap
        $\Delta$. Their superconducting phases $\varphi_\nu$ are not equal if a
        finite supercurrent flows across the DQD structure but only the
        difference between the superconducting phases,
        $\Delta\varphi=\varphi_L-\varphi_R$, is a gauge-invariant quantity,
        which enters the physical observables, and is conveniently absorbed into
        the tunnel couplings to the right lead, whereas the couplings to the
        left lead are strictly real. Furthermore, the behavior of the
        Hamiltonian, Eq.~\eqref{eq-h}, depends drastically on the \emph{sign} of
        the tunnel couplings. This sign is determined by the overlap between the
        wavefunctions in the respective superconducting lead and on the QD.
        Having assumed that the leads are $s$-wave superconductors, the sign
        depends on the orbital parity of the QD levels, i.e., for an even
        orbital both the left and right tunnel coupling have the same sign,
        while they have opposite signs for an odd orbital
        [Fig.~\ref{fig-setup}(b)]. We can recast all the possible sign
        combinations into two possibilities by defining the \emph{total tunnel
        parity}, \cite{Lee14} 
        $\mathcal{P}=\text{sgn}(t_{L1}t_{L2}t_{R1}t_{R2})$, which may be gauged
        arbitrarily into one of the tunnel couplings. Accounting for both of the
        effects, we replace $t_{R1}\rightarrow t_{R1}e^{i\Delta\varphi/2}$ and
        $t_{R2}\rightarrow \mathcal{P} t_{R2}e^{i\Delta\varphi/2}$, where all
        $t_{\nu i}>0$.

        We will discuss the case in which both QDs are in the
        single-level and Coulomb-blockade regime where the level broadening due
        to the tunnel couplings to the leads, $\Gamma_{\nu i}=2\pi
        N(\varepsilon_F)|t_{\nu i}|^2$ with $N(\varepsilon_F)$ the normal-state
        density of states at the Fermi level, is much smaller than the level
        spacing and than the Coulomb repulsion, $\Gamma_{\nu
        i}\ll\delta\varepsilon_i,U_i$. Then the QDs have a well-defined charge
        $(N_1,N_2)$ except close to transport resonances
        [Fig.~\ref{fig-setup}(c)].

        In the presence of interactions, there are several approaches that have
        been used to study transport through a DQD Josephson
        junction, including mean field, \cite{Rozhkov1999a} slave-boson 
        mean field, \cite{Bergeret2006a, Lopez2007a} renormalization group 
        methods, \cite{Krishna1980a, Choi2004a} real-time diagrammatic
        expansion, \cite{Governale2008a} quantum Monte Carlo, \cite{Siano2004a}
        and finite order perturbation theory. \cite{Aashish2000a} Here we
        combine a zero-bandwidth approximation and conventional perturbation
        theory to obtain both reliable results and physical insight. We restrict
        this analysis to the critical Josephson current $I_c$ at low
        temperatures $T$ by setting $T=0$. Even though this is a macroscopic and
        thus directly accessible quantity, we will see that it contains various
        clear signatures of nonlocal Cooper-pair transport, some of which have
        already been observed. \cite{Deacon15} Having obtained the ground-state
        energy $E_0(\Delta\varphi)$ of $H$ either by perturbation theory or by
        exact diagonalization, the critical current is immediately given as the
        maximum supercurrent
        \begin{equation}
            I_c=\frac{2e}{\hbar}
                \max\limits_{\Delta\varphi}
                    \frac{\partial E_0(\Delta\varphi)}{\partial\Delta\varphi}
        \end{equation}
        supported by the ground state of the system.

        \subsection{\label{sec-zbw}Zero-bandwidth approximation}
            We follow Ref.~\citenum{Affleck2000a} and integrate out the
            superconducting leads, yielding the effective lead Hamiltonian
            \begin{align}
                H_{\nu}^\text{zbw}=\Delta_b 
                    c^\dagger_{\nu\uparrow}c^\dagger_{\nu\downarrow}
                    +\text{H.c.}
                \label{eq-hzbw}
            \end{align}
            and renormalized tunnel parameters $t_{\nu
              i}\rightarrow\tilde{t}^{b}_{\nu i}$. Thus, the Josephson
            junction is represented by a four-site superconducting molecule,
            which can be exactly diagonalized. Former studies using the
            zero-bandwidth approach in interacting Josephson junctions have
            proven to show qualitatively good agreement with mean-field
            calculations. This approach retains the essential features of the
            competition between pairing correlations and Kondo correlations
            occurring in the single-QD case. \cite{Vecino2003a,Bergeret2007a}
            The renormalized parameters $\Delta_b$, $t^{b}_{\nu i}$ may be
            obtained from self-consistent calculations. \cite{Affleck2000a} We
            compare the results coming from the zero-bandwidth lead Hamiltonian,
            Eq.~\eqref{eq-hzbw}, with the fourth-order perturbation-theory 
            calculations using the original Hamiltonian, Eq.~\eqref{eq-hlead},
            and find very good agreement already using the bare couplings,
            $\Delta_b=\Delta$ and $t^{b}_{\mu i}$, in the regime where $t_{\nu
            i}\lesssim|\Delta|$, up to a global prefactor, which is proportional
            to the energy density of states of the normal-state leads.
         
        \subsection{\label{sec-pt}Perturbation theory and microscopic behavior} 
            The numerical treatment by exact diagonalization has to be
            accompanied by a perturbative treatment to identify the physical
            processes giving rise to the critical current, in particular, to
            understand which features are attributable to nonlocal Cooper-pair
            transport. Furthermore perturbation theory is not restricted to the
            zero-bandwidth approximation. We calculate the corrections to the
            ground-state energy of the isolated QDs due to their coupling to the
            superconducting leads in fourth order of $t_{i\nu}$, which is the
            leading order of Cooper-pair transfer between the leads. To handle
            the large amount of processes available in fourth order we develop a
            diagrammatic scheme, the details of which are given in
            App.~\ref{app-pt}.

            In the unperturbed ground state, there are no excitations in the
            leads and the QDs have a well-defined charging state $(N_1,N_2)$,
            where $N_i$ is the number of electrons on QD $i$. Each charging
            state is spin degenerate and can be realized by different quantum
            states,
            $\Ket{\alpha,\beta}:=\Ket{\alpha}_\text{QD1}\Ket{\beta}_\text{QD2}$,
            all of which are completely decoupled because the model conserves
            the $z$ projection of the total spin, $S_z$. The only exception are
            the states in the $(1,1)$ sector with $S_z=0$, where degenerate
            perturbation theory in the space spanned by the states
            $\Ket{\uparrow,\downarrow}$ and $\Ket{\downarrow,\uparrow}$ is
            required. Since the total spin is conserved, the degenerate subspace
            is diagonal in the basis of the nonlocal singlet and the nonlocal
            triplet,
            \begin{align}
                |S\rangle&=\frac{1}{\sqrt{2}}\Big(|\up,\down\rangle
                    -|\down,\up\rangle\Big) \\
                |T\rangle&=\frac{1}{\sqrt{2}}\Big(|\up,\down\rangle
                    +|\down,\up\rangle\Big).
            \end{align}
            Spin-exchange processes, i.e., processes which swap the spins of the
            QDs, split the singlet and the triplet. Note that the spin triplets
            with $S_z=\pm1$ behave equivalently to $\Ket{T}$ by spin-rotation
            invariance.

            The terms in the perturbative expansion are \emph{local} or
            \emph{nonlocal}, where local processes involve only one of the QDs 
            whereas nonlocal processes involve both QDs. Furthermore, we call
            all processes \emph{Josephson} processes, in which entire Cooper
            pairs are removed from or added to the superconducting leads due to
            two single-particle tunnel events. In processes which are not
            Josephson processes as many carriers are added to each lead as are
            removed, so we call them \emph{cotunneling} processes. \footnote{As 
                the charging state of the QDs must not change, Josephson
                processes and cotunneling processes do not mix to fourth order, 
                cf.~App.~\ref{app-pt}} 

            Summarizing all of the processes, we can write down the general form
            of the correction of the ground-state energy in perturbation theory,
            \begin{multline}
                \delta E_0(\Delta\varphi)=E^{\loc}_\text{CT,1}
                    +E^{\loc}_\text{CT,2}
                    +E^{\loc}_{J,1}(\Delta\varphi)
                    +E^{\loc }_{J,2}(\Delta\varphi) \\
                    +E^\nonloc_\text{CT,sc}+E^\nonloc_{J,\text{sc}}
                        (\Delta\varphi)
                    \pm\big[E^\nonloc_\text{CT,se}
                    +E^\nonloc_{J,\text{se}}(\Delta\varphi)\big],
                \label{eq-epert}
            \end{multline}
            where the superscript denotes whether the correction is due to local
            (loc) or nonlocal (nl) processes and the subscripts denote whether
            the correction comes from a cotunneling (CT) or from a Josephson (J)
            process. For local processes, the second subscript denotes the QD
            which is involved in the process whereas for nonlocal processes, the
            second subscript denotes whether the process is a spin-exchanging
            (se) or spin-conserving (sc) process. The spin-exchange
            contributions are nonzero only in the $(1,1)$ sector. In the $(1,1)$
            sector, Eq.~\eqref{eq-epert} is thus the energy correction of the
            nonlocal triplet (upper sign) and the nonlocal singlet (lower sign).

            \begin{figure}
                \includegraphics{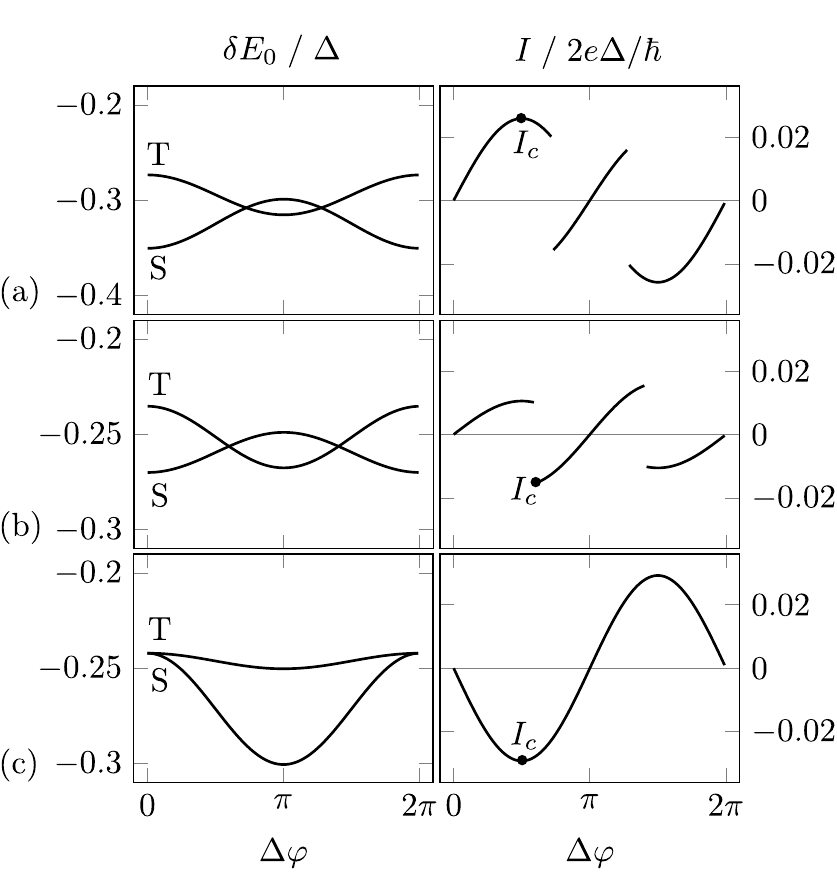}
                \caption{Phase dependence of the energy corrections $\delta E_n$
                    of the two lowest-lying states in the $(1,1)$-charge sector 
                    and the resulting supercurrent $I$ in the ground state
                    evaluated in the zero-bandwidth approximation.
                    (a) Point A in
                    Fig.~\ref{fig-ic}. The critical current $I_c$ is carried by
                    the singlet ground state (S) at $\Delta\varphi=\pm\pi/2$.
                    (b) Point B in Fig.~\ref{fig-ic}. The critical current
                    is carried by the triplet ground state (T) at
                    $\Delta\varphi\neq\pm\pi/2$. 
                    (c) Point C in Fig.~\ref{fig-ic}. At odd total tunnel 
                    parity, $\mathcal{P}=-1$, the ground state is a singlet at
                    all $\Delta\varphi$ and the critical current has the
                    conventional sinusoidal dependence on $\Delta\varphi$. The
                    junction is in the $\pi$ phase.}
                \label{fig-interplay}
            \end{figure}

            In the charge sectors with a unique ground state, the critical
            current is given directly by the amplitude of the phase-dependent
            corrections of the ground-state energy, which, in perturbation
            theory, are proportional to $\cos(\Delta\varphi)$. The amplitude is
            commonly referred to as the (phase-independent) Josephson energy
            $E_J$ and the critical current is proportional to the Josephson
            energy, $I_c\propto E_J$.  Since $E_J$ decomposes into local and
            nonlocal contributions, so does $I_c$. The critical phase is always
            at $\Delta\varphi=\pm\pi/2$, where $\partial_{\Delta\varphi}
            \cos(\Delta\varphi)$ is maximized.  

            In the $(1,1)$-charge sector, the situation is more complicated.
            Both the energy of the singlet state and the energy of the triplet
            state, cf.~Eq.~\eqref{eq-epert}, depend on the phase difference such
            that they may cross for suitable parameters and hence the ground
            state changes between singlet and triplet as a function of the phase
            difference $\Delta\varphi$. Three possible situations are shown in
            Fig.~\ref{fig-interplay}. If there is a singlet--triplet
            ground-state transition
            as a function of the phase difference, the cosinelike energy--phase
            relation of the ground state $\delta E_0(\Delta\varphi)$ becomes a
            piecewise function of the phase difference with two different
            amplitudes and with two different constant energy offsets for the
            singlet state and the triplet state, cf.\ right panels of
            Figs.~\ref{fig-interplay}(a) and (b). When the critical current is
            probed, the junction adjusts to the phase difference which maximizes
            the supercurrent. This is not necessarily at the conventional value
            $\Delta\varphi=\pm\pi/2$. At $\Delta\varphi=\pm\pi/2$ there might be
            a singlet (triplet) ground state with a low amplitude
            $E_J(\Delta\varphi)$ which cannot carry as high a supercurrent as
            the triplet (singlet) ground state at a different phase difference
            $\Delta\varphi'\neq\pm\pi/2$ but with a larger amplitude
            $E_J(\Delta\varphi')$ such that $|E_J(\pm\pi/2)|<
            |E_J(\Delta\varphi')\sin(\Delta\varphi')|$
            [Fig.~\ref{fig-interplay}(b)]. \footnote{An important reason that 
                the critical current differs between the singlet phase and the
                triplet phase is the sign of the nonlocal Josephson current. As
                can be seen from Eqs.~\eqref{eq-epert} and~\eqref{eq-lawcos}, it
                depends on the phase, on the parity, and on whether the ground
                state is a singlet or a triplet. So if the local supercurrents
                and the nonlocal supercurrent are flowing in opposite directions
                at $\Delta\varphi=\pm\pi/2$, it can be beneficial to switch to
                the other ground state at $\Delta\varphi'\neq\pm\pi/2$, where
                the individual supercurrents are smaller but flow in the same
                direction.  Due to this interplay it is nontrivial to isolate
                nonlocal features from the critical current.}
            Then the junction will switch to the triplet (singlet) ground state
            and the critical phase locks to $\Delta\varphi'$.

    \section{Results}
        \begin{figure*}
            \includegraphics{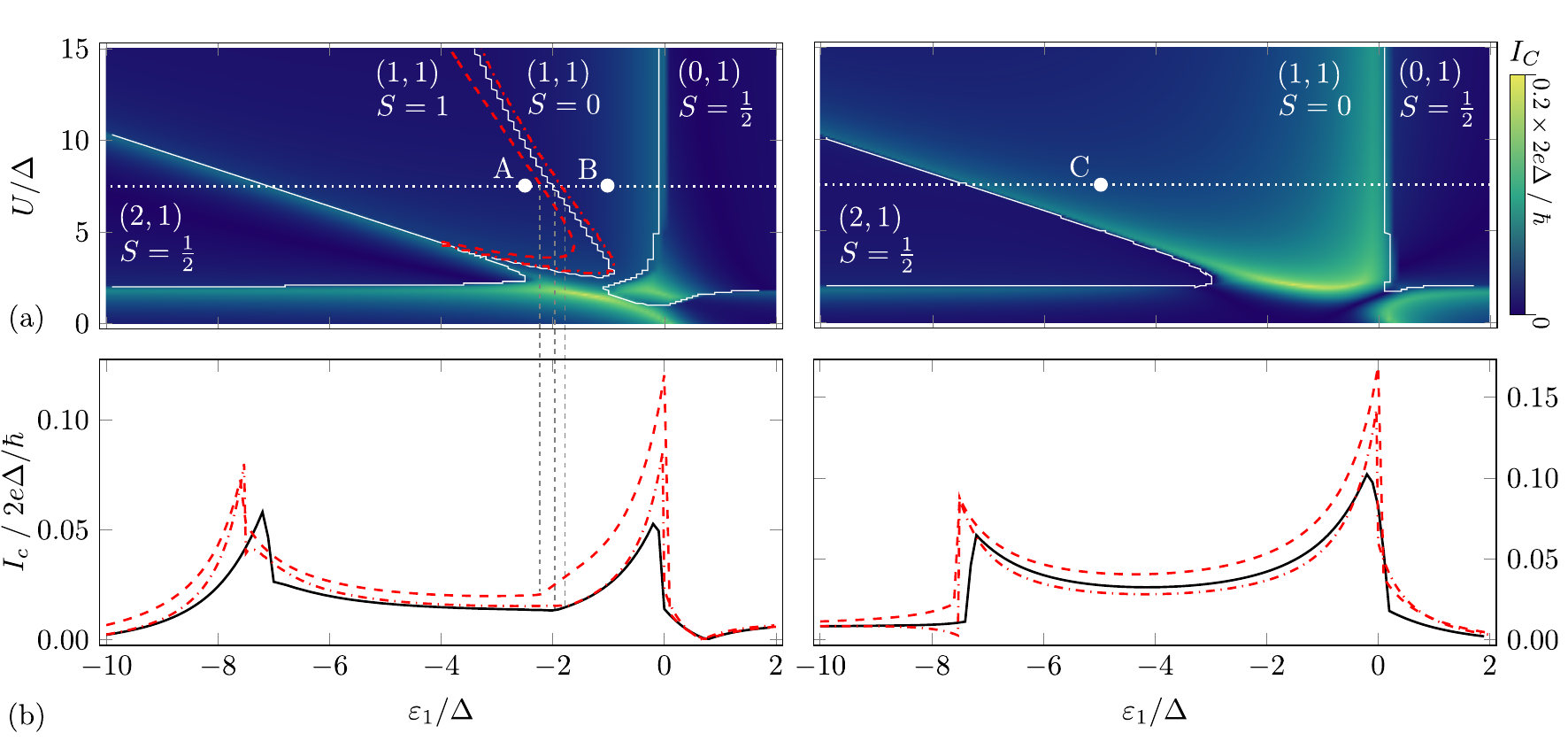}
            \caption{(Color online) (a) Critical current and total spin of the 
                Josephson junction depending on the on-site energy
                $\varepsilon_1$ on quantum dot 1 and on the Coulomb repulsion
                $U=U_1=U_2$. Quantum dot 2 is kept at
                $\varepsilon_2=-1.5\Delta$ and the tunnel couplings are $|t_{\nu
                i}|=0.5\Delta$. Following the white dotted line from
                left to right, the quantum-dot occupation varies
                $(2,1)\rightarrow(1,1)\rightarrow(0,1)$. The critical current
                increases at each transition because the particle number
                fluctuates. Left: at even total tunnel parity, $\mathcal{P}=1$,
                an additional ground-state transition between a nonlocal singlet
                and a nonlocal triplet occurs in the $(1,1)$ sector. It is
                caused by competing cotunneling processes between the quantum
                dots via the superconducting leads which give rise to an
                exchange interaction (text). The red lines indicate the phase
                boundary obtained in perturbation theory in the zero-bandwidth
                approximation (dash dotted) and in the wide-band limit (dashed).
                Right: at odd total tunnel parity, $\mathcal{P}=-1$, the
                singlet--triplet transition is absent. 
                (b) Cuts across the $\varepsilon_1$--$U$ plane at $U=7.5\Delta$
                reveal that the shape of the current peaks depends strongly on
                the tunnel parity. This can be traced back to the
                singlet--triplet transition (text), which also immediately
                manifests as a kink in the critical current. Since singlet and
                triplet can be distinguished only by nonlocal transport, this
                kink is immediate evidence of coherently split Cooper pairs.
                There is no qualitative difference between the zero-bandwidth
                approximation (exact: solid black, perturbative: dash-dotted
                red) and the wide-band limit (dashed red). In the wide-band
                limit, the critical current scales with the density of states,
                which is chosen to agree with the result in zero-bandwidth
                approximation.}
            \label{fig-ic}
        \end{figure*}

        \subsection{Singlet--triplet ground-state transition}
            In Fig.~\ref{fig-ic}, we present the results for the critical
            current of the DQD Josephson junction. In Fig.~\ref{fig-ic}(a), we
            plot the critical current and the total spin of the QD system in the
            ground state carrying the critical current depending on $U=U_1=U_2$
            and on $\varepsilon_1$ at fixed $\varepsilon_2=-1.5\Delta$ in the
            zero-bandwidth limit.  When the tunnel parity is even,
            $\mathcal{P}=1$, (left panels) the total spin in the $(1,1)$ sector
            changes from a singlet to a triplet in a regime of finite charging
            energy $U$. This is true both in the zero-bandwidth limit (solid
            and dash-dotted phase boundaries) and in the wide-band limit (dashed
            phase boundary) with no qualitative differences. In the wide-band
            limit, the normal-state density of states in the superconducting
            leads is constant at all relevant QD energies. Then it affects the
            critical current only as a constant prefactor, which we choose to
            fit the zero-bandwidth results.  With the tunnel couplings $|t_{\nu
            i}|=0.5\Delta$, as chosen in Fig.~\ref{fig-ic}, and a
            superconducting gap on the order of $\Delta=0.1$~meV, we obtain a
            critical current of a few nanoampere at the resonances.  This agrees
            with the experimental data of Ref.~\citenum{Deacon15}, where
            aluminum electrodes were used.
            \cite{Deacon15} \footnote{Substituting the ground-state energy with
                the free energy, $I=-2ekT/\hbar\times
                \partial_{\Delta\varphi}\ln\sum_n e^{-\beta E_n}$,
                \cite{Bruus04} we verify that the total-spin transition is
                present also at finite temperatures, $kT\sim0.1\Delta$, and
                hence experimentally accessible.}

            In the wide-band limit at very large Coulomb repulsion,
            $U_i\rightarrow\infty$, the parameter space is confined to what is
            the upper right corner of the $(1,1)$ sector in
            Fig.~\ref{fig-setup}(c) and the triplet ground state cannot be
            observed, consistent with earlier studies. \cite{Choi00}
            Numerically, we confirm that the triplet ground state can emerge if
            either the Coulomb repulsion or the bandwidth are not significantly
            larger than all other energy scales, which makes it rather the rule
            than the exception.

            Intuitively, one could expect a singlet ground state in the
            $(1,1)$-charge sector, which could be justified by noting that only
            the nonlocal singlet can tunnel into the leads to hybridize with the
            Cooper pairs in the $s$-wave superconductors and lower its energy.
            However, in addition to this second-order \emph{Cooper-pair}
            tunneling process, there are \emph{genuine} fourth-order terms with
            additional intermediate single-particle excitations in the
            superconducting leads. They cannot be decomposed into two
            Cooper-pair tunnel events and may favor the triplet ground state.

            The splitting of the nonlocal singlet and the nonlocal triplet is
            given by $E^\nonloc_{CT,se}+E^\nonloc_{J,se}(\Delta\varphi)$.  Since
            the superconductors are identical, the energy corrections can be
            split into a matrix element
            $\mathcal{M}^{\nonloc/\loc}_{\mathrm{CT/J,se/sc}}$ and the tunnel
            couplings to the leads. Since we associate the parity $\mathcal{P}$
            and the phase difference $\Delta\varphi$ to the tunnel couplings
            $t_{\nu i}$, the matrix element is then independent of either. Each
            nonlocal process contributing to
            $E^\text{nl}_{J,\text{se}}(\Delta\varphi)$ can involve the same
            superconducting lead twice or both leads once. Summing all
            combinations we obtain
            \begin{multline}
                E^\text{nl}_{J,\text{se}}(\Delta\varphi)=
                    \mathcal{M}^\text{nl}_{J,\text{se}} \\
                    \times\Big[
                (t_{R1}t_{R2})^2+(t_{L1}t_{L2})^2 \\
                    +2\mathcal{P}t_{R1}t_{R2}t_{L1}t_{L2}\cos(\Delta\varphi)
                    \Big],
                \label{eq-lawcos}
            \end{multline}
            where we find $\mathcal{M}^\text{nl}_{J,\text{se}}>0$, such that
            $E^\text{nl}_{J,\text{se}}$ is strictly positive and favors the
            singlet ground state. The details of the calculation of the matrix
            elements are given in App.~\ref{app-pt}.

\begin{figure}
                \includegraphics{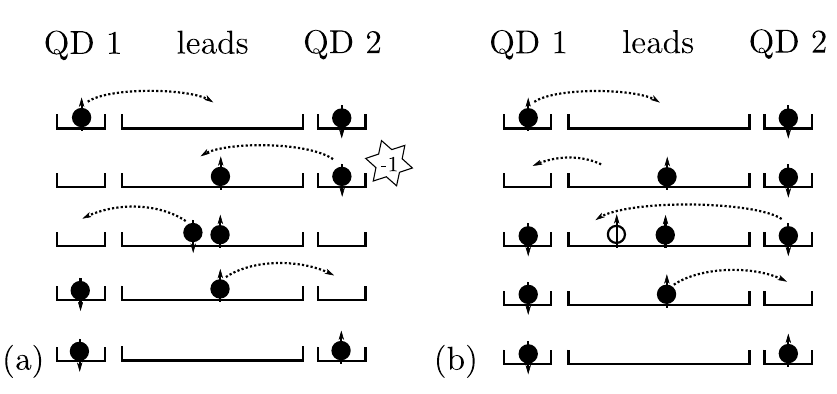}
                \caption{Two spin-exchange processes which have a different
                    overall sign and hence energetically favor (a) singlet
                    states and (b) triplet states.  Initially, one of two
                    electrons (filled circles) with opposite spin resides on
                    each quantum dot (left and right narrow tray). A final
                    state with the spins swapped can be reached via
                    intermediate virtual states (arranged top to bottom)
                    connected by four tunnel processes (dashed arrows)
                    between the quantum dots and the superconducting leads
                    (wide tray). Every time the left-to-right order of two
                    fermions is changed, a sign results.  (a) If only
                    electronlike states in the leads are involved, the two
                    initial electrons have to be swapped.  This kind of
                    process with a negative sign energetically favors the
                    singlet state.  (b) If the exchange process involves a
                    hole (empty circle), it is possible to exchange the
                    spins without anticommutation signs. This type of
                    process energetically favors the triplet state.}
                \label{fig-sign}
            \end{figure}

            With the same arguments, we find
            \begin{multline}
                E^\text{nl}_\text{CT,se}=\mathcal{M}^\text{nl}_\text{CT,se} 
                \\ \times
                    \Big[(t_{R1}t_{R2})^2+(t_{L1}t_{L2})^2
                        +2\mathcal{P}t_{R1}t_{R2}t_{L1}t_{L2}\Big]\\
                =\mathcal{M}^\text{nl}_\text{CT,se}
                    (t_{R1}t_{R2}+\mathcal{P} t_{L1}t_{L2})^2,
                \label{eq-ctse}
            \end{multline}
            so the sign of $E^\text{nl}_\mathrm{CT,se}$ is determined solely by
            $\mathcal{M}^\mathrm{nl}_\mathrm{CT,se}$. This time, however, the
            perturbative analysis of $\mathcal{M}^\text{nl}_\text{CT,se}$
            reveals that processes of both signs exist. Processes in which the
            electrons are exchanged via electronlike excitations in the leads
            [Fig.~\ref{fig-sign}(a)] have a different number of fermion-exchange
            signs than processes in which the electrons are exchanged via an
            electronlike \emph{and} a holelike excitation
            [Fig.~\ref{fig-sign}(b)]. Processes involving only electronlike
            excitations lower the singlet while process involving an
            electronlike and a holelike excitation lower the triplet. All
            spin-exchange processes are listed in App.~\ref{app-pt}.

            For the existence of a triplet ground state the Josephson processes
            play a minor role as they are suppressed if $\Delta\varphi$ is
            chosen such that $\mathcal{P}\cos\Delta\varphi=-1$. The nonlocal
            triplet is thus driven by the sign of $E^\nonloc_{\mathrm{CT,se}}$,
            which is ultimately determined by the microscopic parameters.

            The influence of $\Delta$ on the parameter space in which there may
            be a triplet ground state can be estimated. The matrix element of
            each process in the perturbative expansion is weighted by the
            product of the reciprocal virtual excitation energies
            (cf.~App.~\ref{app-pt}).  In electronlike processes, which favor the
            singlet ground state, all virtual states involve excitations on the
            QDs. They can be estimated by
            \begin{equation}
                \frac{1}{(\varepsilon_\mathrm{DQD}+\Delta)^2}
                \frac{1}{2\varepsilon_\mathrm{DQD}+2\Delta},
            \end{equation}
            where $\varepsilon_{\mathrm{DQD}}$ is a typical
            DQD-excitation energy. By using electronlike and holelike
            excitations, however, it is possible to restore the initial DQD
            state at the expense of two virtual excitations in the leads.  These
            processes, which favor the triplet ground state, are hence weighted
            by
            \begin{equation}
               \frac{1}{(\varepsilon_\mathrm{DQD}+\Delta)^2}
                   \frac{1}{2\Delta}.
            \end{equation}
            If $\Delta$ is comparable or smaller than
            $\varepsilon_{\mathrm{DQD}}$, the ratio between triplet-favoring and
            singlet-favoring processes, $1+\varepsilon_\text{DQD}/\Delta$, 
            becomes large and a triplet ground state may emerge. 

            In general, second-order Cooper-pair tunneling restores the ground
            state of the superconducting leads in one intermediate virtual
            state, whereas the leads are excited in all three intermediate
            states of genuine fourth-order processes. So, genuine fourth-order
            processes have an additional suppression by $\Delta^{-1}$ compared
            to second-order Cooper-pair processes. \footnote{When integrating
                out the momentum quantum number in the case of continuous leads,
                at large $\Delta$, second-order Cooper-pair processes are
                independent of $\Delta$ so the relative suppression of genuine
                fourth-order processes may be even stronger.}
            If the superconducting gap is very large compared to the other
            energy scales, the singlet character induced by the superconducting
            leads dominates and we recover the intuitive singlet ground state.
            We investigate this limit in more detail in Sec.~\ref{sec-infdelta}.  

            The triplet ground state is absent in the regime of odd total tunnel
            parity, $\mathcal{P}=-1$, [Fig.~\ref{fig-ic}(a), right panel]. This
            is because different cotunneling processes interfere destructively,
            which reduces the magnetic exchange coupling. At negative tunnel
            parity, $\mathcal{P}=-1$, the parity-dependent factor in
            Eq.~\eqref{eq-ctse} is reduced and even vanishes in a symmetric
            setup, $t_{Li}=t_{Ri}$. Without the exchange coupling, the nonlocal
            Josephson processes will always favor the singlet over the triplet,
            cf. Eq.~\eqref{eq-lawcos} and Fig.~\ref{fig-interplay}(c).

        \subsection{Peak asymmetry and signature of nonlocal transport}
            \begin{figure}
                \includegraphics{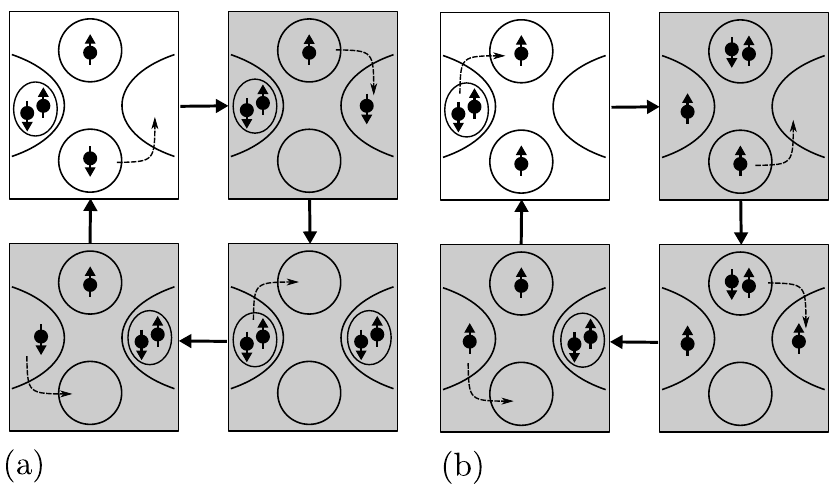}
                \caption{Typical Josephson transport processes via three
                    intermediate virtual states (gray) in the $(1,1)$-charge
                    sector. 
                    (a) In the singlet ground state, there are transport
                    channels in which the two electrons initially localized on
                    the DQD are absorbed as a Cooper pair in one lead. 
                    (b) In the triplet ground state (all triplets are equivalent
                    by spin-rotation invariance and time-reversal symmetry), the
                    electrons of the Cooper pair need to be transferred
                    sequentially through the double quantum dot.}
                \label{fig-transport}
            \end{figure}

            Fig.~\ref{fig-ic}(b) shows the critical current as a function of
            $\varepsilon_1$ for a fixed on-site repulsion $U=7.5\Delta$ in the
            $\mathcal{P}=\pm1$ regimes. Red lines are the results from
            perturbation theory in the zero-bandwidth limit (dash dotted) and in
            the wide-band limit (dashed), both of which agree with the exact
            results of the zero-bandwidth model (black solid). In general, the
            critical current is high at the charge neutrality points where the
            number of electrons on the QDs can fluctuate. 

            Both the singlet ground state and the triplet ground state can
            support a finite supercurrent. In the singlet phase, the
            supercurrent tends to be higher because there is an additional
            transport channel where the two electrons of a Cooper pair are
            simultaneously added to or removed from the DQD. In the triplet
            ground state, this channel is blocked by the Pauli exclusion
            principle.  At the resonance near $\varepsilon_1=0$ in
            Fig.~\ref{fig-ic}(b), the QD charging states $(1,1)$, $(1,0)$,
            $(0,1)$, and $(0,0)$ are almost degenerate so this type of transport
            is particularly strong and in the singlet ground state the
            supercurrent is primarily carried by the process shown in
            Fig.~\ref{fig-transport}(a). By inspecting all possible
            combinations, it is easy to see that there is no fourth-order
            transport process in the triplet ground state involving the
            $(0,0)$-charging state. \footnote{It is helpful to consider the
                equivalent behavior of the fully-polarized triplet.}
            At the other resonance, however, $\varepsilon_1\approx-U_1$, a
            Josephson process involving the almost-degenerate QD states $(1,1)$,
            $(1,0)$, $(2,1)$, and $(2,0)$ does exist in the triplet ground state
            [Fig.~\ref{fig-transport}(b)]. So with increasing $\varepsilon_1$ in
            Fig.~\ref{fig-ic}(b), the singlet ground state has resonances both
            at the $(2,1)$--$(1,1)$ transition and at the $(1,1)$--$(0,1)$
            transition but the triplet ground state has only one resonance at
            the $(2,1)$--$(1,1)$ transition. Hence, at even parity, with
            increasing $\varepsilon_1$ the critical current decreases in the
            $(1,1)$ sector as long as the system is still in the triplet ground
            state. Only once the ground state switches to a singlet, which
            happens close to the $(1,1)$--$(0,1)$ transition
            [cf.~Fig.~\ref{fig-ic}(a)], the critical current rises again,
            producing a notable asymmetry between the resonance peaks. At odd
            parity, there is no asymmetry because the ground state remains a
            singlet throughout the entire $(1,1)$ sector.

            We emphasize that the singlet--triplet transition of the ground
            state in the $(1,1)$ sector, realized in a large parameter window,
            leads to a \emph{kink} in the critical current as a function of
            $\varepsilon_2$.  This kink appears because, in the singlet phase,
            different processes contribute to the critical current than in the
            triplet phase and, hence, the dependency on the on-site energies
            changes across the singlet--triplet transition. Since the
            distinction between triplet and singlet phases results from phase
            coherent and \emph{nonlocal} exchange, its observation in the
            critical current is a clear sign of nonlocal Cooper-pair transport.

            At odd tunnel parity, $\mathcal{P}=-1$, there is no singlet--triplet
            transition and hence no signature of nonlocal transport in the
            critical current. The other way around, if two neighboring resonance
            peaks belonging to the same level of a QD decay symmetrically in the
            offresonant regime between them, the level has odd parity. An
            asymmetric decay may be caused by a singlet--triplet transition and
            indicates even parity.

    \section{Multilevel quantum dot}               
        \begin{figure}
            \includegraphics{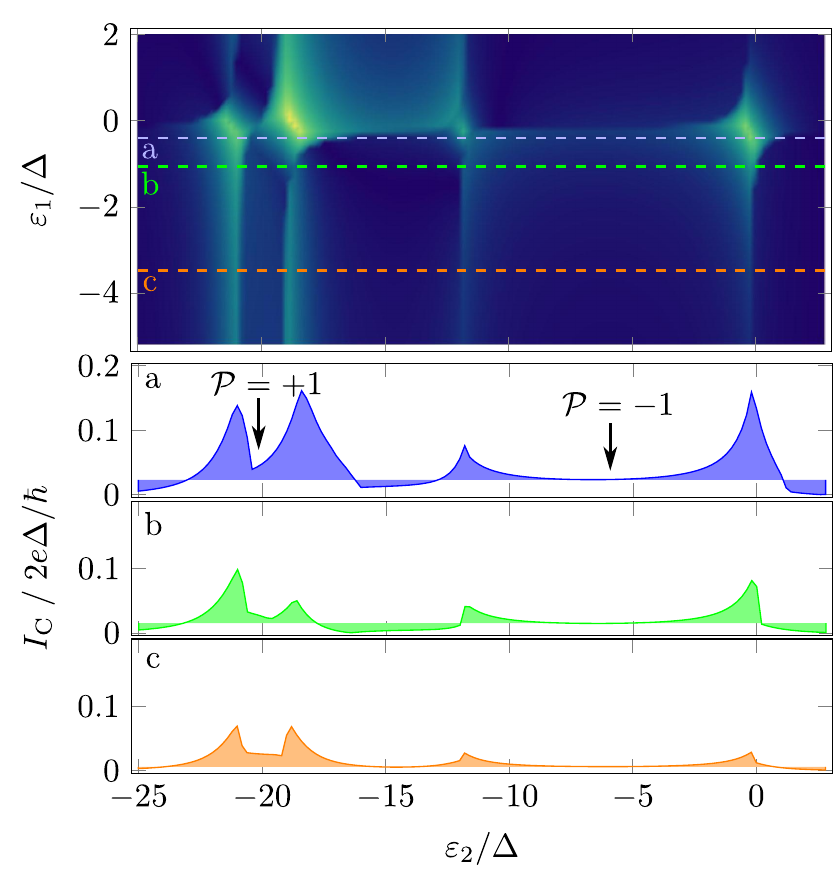}
            \caption{(Color online) Critical current of the double-quantum-dot  
                junction with two different levels of opposite parity on quantum
                dot 1.  The parameters of the quantum dots are
                $\delta=18.5\Delta$, $t_{R1}=t_{L1}=0.45\Delta$,
                $t_{R21}=-t_{L21}=0.45\Delta$, $t_{R22}=t_{L22}=0.57\Delta$,
                $U_1=28\Delta$, $U_{11}=12\Delta$, $U_{22}=3\Delta$, and
                $U_{12}=0.5\Delta$.  Upper panel: critical current as a function
                of the gate-controlled on-site energies $\varepsilon_1$ and
                $\varepsilon_2$.  Lower panel: cuts at (a)
                $\varepsilon_1=-0.9\Delta$, (b) $\varepsilon_1=-1.8\Delta$, and
                (c) $\varepsilon_1=-4.2\Delta$. In the absence of nonlocal
                transport, the three curves are expected to differ only by a
                constant. Instead, when approaching the resonance,
                $\varepsilon_1\rightarrow0$, the critical current grows more
                strongly at the peaks and less strongly between the peaks.
                Within our model this behavior is clearly attributable to
                nonlocal coherent transport and it was already observed 
                experimentally. \cite{Deacon15}}
            \label{fig-deacon}
        \end{figure}

        In order to make contact to the experiments presented in
        Ref.~\citenum{Deacon15}, where multiple QD levels were probed, we
        include one extra level in the model, e.g., on QD~2. In this way, we can
        study the evolution of the critical current along four consecutive
        resonances by continuously tuning $\varepsilon_2$.  This scenario
        requires the substitution of $H_2$ in Eq.~\eqref{eq-h} by
        \begin{multline}
            H_2=\varepsilon_2d^\dagger_{21\sigma}d_{21\sigma}
                +(\varepsilon_2+\delta) d^\dagger_{22\sigma}d_{22\sigma}\\
                +\sum_{(i\sigma)\neq(j\rho)}U_{ij}n_{2i\sigma}n_{2j\rho},
        \end{multline}
        where $\delta$ is the energy separation between the QD levels, and
        $U_{ij}$ the Coulomb energy coming from the interaction of the
        occupation of the levels $i$ and $j$ on the second QD. \footnote{Here, 
            we neglect the spin-exchange interaction within QD~2. Adding it,
            however, would not change our results because the device is in the
            single-level regime, $\delta\gg U_{2i},t$.} 
        Besides, we also need to include an additional tunnel coupling to
        Eq.~\eqref{eq-ht}.  Computationally, the addition of the extra level
        requires to extend the $256\times256$ Hamiltonian matrix to a
        $1024\times1024$ matrix, which remains tractable. Taking into account
        that the levels are well separated, we can still define the total tunnel
        parity close to a resonance as within the single-level model involving
        only the four relevant tunnel couplings. 

        Choosing the measurement presented in Fig.~4 of Ref.~\citenum{Deacon15}
        as a specific example, we observe that two neighboring resonance peaks
        at lower gate voltages (higher on-site energies) are clearly more
        symmetric than two neighboring resonances at higher gate voltages (lower
        on-site energies).  Within our model this is expected if the two lower
        peaks belong to one level with odd parity and the two higher peaks
        belong to one level with even parity (cf.~Fig.~\ref{fig-ic}). Note that
        concerning the occupation numbers this does not agree with
        Ref.~\citenum{Deacon15}, which seems to suggest that, in total, three
        levels on QD~2 are involved.  Nevertheless, the model is clearly capable
        of reproducing the qualitative features observed in the experiment when
        choosing the appropriate parameters. 

        In Fig.~\ref{fig-deacon} we show the critical current as a function of
        $\varepsilon_2$ and $\varepsilon_1$ (top panel) and in the lower panel
        we perform three cuts at different values of $\varepsilon_1$.  Close to
        the resonance (blue and green curves), the results are basically
        equivalent to the results from the single-level model, once with even
        parity, and once with odd parity. Here, we recover the signature of
        nonlocal transport proposed in Ref.~\citenum{Deacon15}: if there were
        only two independent transport channels, local transport through QD~1
        and local transport through QD~2, the blue and the green curve would
        only differ from each other by being shifted along the vertical axis.
        This is because changing $\varepsilon_1$ would only affect the
        contribution of the critical current going through QD~1, which is
        independent from $\varepsilon_2$, i.e., it cannot influence the behavior
        of the critical current along the horizontal axis in the lower panel of
        Fig.~\ref{fig-deacon}. Choosing, however, an arbitrary reference point
        as indicated by the shaded areas, we can clearly see that there
        \emph{is} cross-talk between $\varepsilon_1$ and $\varepsilon_2$. When
        QD~1 is brought closer to resonance, the resonance peaks of QD~2 grow,
        indicating an additional transport channel involving both QD~1 and QD~2.
        Moreover, there are interference effects which \emph{reduce} the
        critical current between the two levels on QD~2, when QD~1 is brought
        closer to resonance. More strikingly, for values of 
        $-\varepsilon_1>t,U$, we observe that the resonance at
        $\varepsilon_2\approx-18\Delta$, \emph{increases} when effectively
        decoupling QD~1 (yellow curve).  Now, the Cooper pairs tunnel locally
        through QD~2 but through two different levels. Note that this feature
        cannot occur in the simpler model with only two single-level QDs.
        Summing up, our model reproduces the signatures of nonlocal transport
        observed in Ref.~\citenum{Deacon15} even though, as we have argued in
        Sec.~\ref{sec-pt}, the actual decomposition of the Josephson energy is
        more complicated than stated in their work.

    \section{Regime of dominant nonlocal transport}
        \label{sec-infdelta}
        \begin{figure}
            \includegraphics{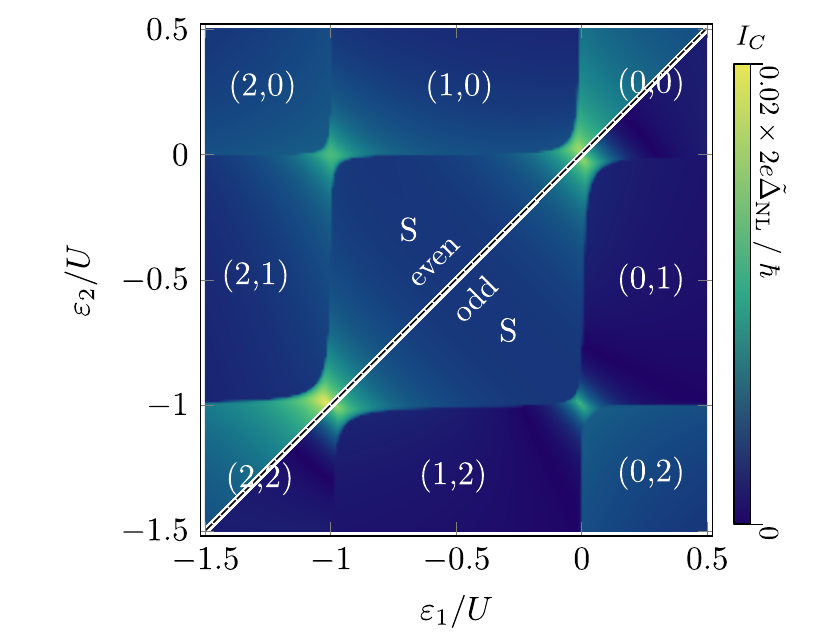}
            \caption{(Color online) Critical current in the limit of large
                $\Delta_\nu$, where $U_1=U_2\equiv U$,
                $\tilde\Delta_\text{L}=0.05U$, $\tilde\Delta_\text{NL}=0.025U$,
                and $\tilde t=0.01U$. The plot for even and the plot for odd
                parity are separated by a dashed line.  Electrons can leave or
                enter the superconducting leads only in pairs, so sequential
                transport is not possible and the single-particle resonances at
                the ground-state transitions are suppressed. In turn, if the
                Coulomb repulsion is large, two-particle resonances in the
                ground state are possible only at four points, where
                $\varepsilon_i=0$ or $\varepsilon_i=-U_i$ is fulfilled
                simultaneously for both quantum dots $i=1,2$. At these points,
                nonlocal transport dominates. Parity has only quantitative
                influence. In particular, the singlet ground state is stable.}
            \label{fig-infdelta}
        \end{figure}

        Finally, we comment shortly on the regime in which nonlocal transport is
        dominant, similar to the Cooper-pair-splitter regime of
        superconductor--normal junctions proposed in Ref.~\citenum{Recher01}. In
        this regime, the superconducting gap is significantly larger than the
        on-site energies and than the tunnel couplings,
        $\Delta_\nu\gg\varepsilon_j,t_{\mu j}$ and excitations in the
        superconducting leads are highly unfavorable. This suppresses both the
        cotunneling contributions which allow for a triplet ground state and the
        sequential Josephson processes which enable a supercurrent to flow in
        the triplet ground state. As a consequence, there is never a triplet
        ground state in the limit of large $\Delta_\nu$.

        When the QDs are brought close to the $(0,0)$--$(1,1)$ point or to the
        $(1,1)$--$(2,2)$ point in the stability diagram, the only remaining
        transport channel is nonlocal, i.e., the Cooper pairs have to be split:
        simultaneous transport of a pair through a single QD is suppressed by
        the Coulomb repulsion $U_i$ and sequential transport of single electrons
        originating from the same Cooper pair through the QDs is suppressed by
        the superconducting gap $\Delta_\nu$. In this situation, the model can
        be simplified further by completely integrating out the superconducting
        leads \cite{Recher10, Schroer15} to leading order in $t_{\nu
        i}/\Delta_\nu$.  Assuming that $|\varepsilon_i|,U_i,t_{\nu
        i}\ll\Delta_\nu$ and absorbing a renormalization of the on-site energies
        into $\varepsilon_i$, we obtain 
        \begin{align} 
            H_\text{eff}&=\sum_{i\sigma}\varepsilon_i
                d_{i\sigma}^\dagger d_{i\sigma} 
                +\sum_i U_i d_{i\uparrow}^\dagger
                    d_{i\downarrow}^\dagger d_{i\uparrow} d_{i\downarrow} 
            \notag\\
            &+\sum_{ij\sigma\nu}\tilde\Delta_{ij}\mathcal{P}_\nu 
                e^{-i\varphi_\nu} 
                \sigma d_{i\sigma}^\dagger d_{j\bar\sigma}^\dagger +\text{H.c.} 
            \notag\\
            &+\sum_{ij\sigma\nu} \tilde t\mathcal{P}_\nu 
                d_{i\sigma}^\dagger d_{j\sigma}, 
            \label{eq-hinfdelta} 
        \end{align} where $\tilde\Delta_{ij}$
        is the effective amplitude to inject a local ($i=j$) or a nonlocal
        ($i\neq j$) Cooper pair and $\tilde t$ describes cotunneling. The parity
        enters via $\mathcal{P}_\nu=1$ if $\nu=L$ and
        $\mathcal{P}_\nu=\mathcal{P}$ if $\nu=R$. Because there are no
        sequential Cooper-pair transport processes as shown in
        Fig.~\ref{fig-transport}(b), the triplet sector cannot support any
        supercurrent and decouples completely. We obtain the critical current in
        the limit of large $\Delta_\nu$ by exact diagonalization of
        Eq.~\eqref{eq-hinfdelta} (Fig.~\ref{fig-infdelta}). 

        At a first glance, the behavior of the critical current looks
        deceptively similar to the results discussed before with resonancelike
        features along the ground-state transitions. But sequential transport is
        not possible without intermediate excitations in the superconducting
        leads. So in the $(1,1)$ ground state, transport is possible only in
        resonance with the $(0,0)$ state, which requires
        $\varepsilon_1+\varepsilon_2=0$, and in resonance with the $(2,2)$
        state, which requires $\varepsilon_1+\varepsilon_2=
        2(\varepsilon_1+\varepsilon_2)+U_1+U_2$. The ground state, however, is
        $(1,1)$ only if $\varepsilon_i<0$ and $\varepsilon_i<
        2\varepsilon_i+U_i$. This rules out all configurations except
        $\varepsilon_1=\varepsilon_2=0$ and $\varepsilon_i=-U_i$. Two more
        nonlocal resonances are found similarly in the sector of odd total
        occupation when $\varepsilon_{1,2}=0$ and simultaneously
        $\varepsilon_{2,1}=-U_{2,1}$. 
       
        Local transport, on the other hand, is resonant if the $(N_1,0)$ state
        is in resonance with the $(N_1,2)$ state or if the $(0,N_2)$ state is
        resonant with the $(2,N_2)$ state. In any case, $0=2\varepsilon_i+U_i$
        is required. But because $U_i>0$, this condition is incompatible with
        any ground state in which QD $i$ is either empty or doubly occupied.
        Hence there is only nonlocal supercurrent carried by the ground state.
        Note that the argument is valid only if the resonances are sharp
        compared to the level spacing, i.e., if the Coulomb repulsion $U_i$ is
        sufficiently large.
        
        This gives a relatively straight-forward signature of nonlocal
        transport: localized resonances in the $\varepsilon_1$--$\varepsilon_2$
        plane indicate nonlocal transport. If, on the other hand, only extended
        steplike features are visible, they are most likely attributable to
        ground-state transitions and there is no nonlocal Cooper-pair transport.

    \section{Conclusion}
        In this work, we have considered the setup of a double-quantum-dot
        Josephson junction. We have used exact diagonalization and perturbation
        theory in the tunneling from the quantum dots to the superconducting
        leads in order to calculate the critical current of the junction. We
        included all possible occupations of the quantum-dot levels for various
        values of the quantum-dot level energies including finite on-site
        Coulomb repulsion. Depending on the parity of the quantum-dot levels, we
        discovered a nonlocal (one electron per quantum dot) singlet--triplet
        ground-state transition (at total even parity) as a function of the
        quantum-dot level energies for a large parameter window when the
        superconducting gaps are smaller than or comparable to the quantum-dot
        energy scale. This transition becomes visible as a kink in the critical
        current with an associated asymmetric line shape between resonances,
        which could serve as a new sign of coherent Cooper-pair splitting. We
        consistently find this physics in zero-bandwidth approximation, where
        each superconducting lead is modeled as a single site with pairing
        interaction, as well as in the wide-band limit of continuous open leads.
        Regarding recent experiments \cite{Deacon15} on this setup, we included
        an additional level with opposite parity on one of the quantum dots. We
        observed critical current traces by varying the gate voltage of the
        quantum dot with two levels for different but fixed level energies of
        the other quantum dot that are consistent with the experiment: the
        different traces are not just shifted by a constant offset but show
        enhanced relative current profiles near resonances, which, in our model,
        is directly related to coherent Cooper-pair transport via different
        quantum dots as was conjectured in Ref.~\citenum{Deacon15}. In addition,
        we observe asymmetric line shapes between a pair of subsequent
        resonances due to nonlocal singlet--triplet ground-state transitions
        associated with quantum-dot levels showing even total parity. Such
        asymmetries are also visible in the experiment.  Finally, we analyze the
        model in the limit of large superconducting gaps, where we can integrate
        out the superconductors thereby creating a proximity effect in the
        quantum dots.  The resulting effective model for the quantum dots can be
        solved exactly and we find that all current resonances are dominated by
        nonlocal processes in this limit.

    \begin{acknowledgements}
        We gratefully acknowledge discussions with M.-S. Choi and R. S. Deacon,
        helpful comments on the manuscript by B.  Trauzettel and financial
        support by the EU-FP7 Project SE2ND, No.~271554, Spanish Mineco through
        grant FIS2014-55486-P, the DFG, Grant No.~RE 2978/1-1 and Research
        Training Group GrK1952/1 ``Metrology for Complex Nanosystems'', and the
        Braunschweig International Graduate School of Metrology B-IGSM. ALY
        acknowledges financial support from the Spanish MINECO, through the
        ``Mar\'ia de Maeztu'' Programme for Units of Excellence in R\&D
        (MDM-2014-0377).
    \end{acknowledgements}

    \begin{appendix}
        \section{\label{app-pt}Diagrammatic perturbation theory}
            The correction of the ground-state energy can be
            obtained conveniently using the Schrieffer--Wolff transformation, 
            which decouples the low-energy degrees of freedom, i.e., in our case
            the (possibly degenerate) ground state $m$, from the high-energy
            degrees of freedom, i.e., all excited states $l$, up to fourth order
            in the tunneling between the QDs and the superconducting leads.
            Following Ref.~\citenum{Winkler2003}, the fourth-order contribution
            to the resulting low-energy Hamiltonian matrix is
            \begin{multline} 
              \label{eq-fourthorderterms}  
              H_{mm'}^{(4)}=
                -\frac{1}{2}\sum_{l,l',m''}H_{ml}H_{lm''}H_{m''l'}H_{l'm'}\\
                \Big(\frac{1}{(E_m-E_l)^2(E_m-E_{l'})}+%\\
                    \frac{1}{(E_m-E_l)(E_m-E_{l'})^2}
                \Big)\\
                +\sum_{l,l',l''}H_{ml}H_{ll'}H_{l'l''}H_{l''m'}\\
                \frac{1}{E_m-E_l}\frac{1}{E_m-E_{l'}}\frac{1}{E_m-E_{l''}},
            \end{multline}
            where $E_\eta$ is the energy of the unperturbed state $\eta$ and
            $H_{ij}$ is the matrix element for transitions between states $i$
            and $j$. If the unperturbed ground state is unique, $H^{(4)}$ is one
            dimensional and equivalent to the ground state energy correction
            $\delta E_0$.  In the degenerate $(1,1)$-charge sector with $S_z=0$,
            $H^{(4)}$ is two dimensional. For the calculation of the matrix
            elements it is convenient to choose as basis
            $|\uparrow,\downarrow\rangle$ and $|\downarrow,\uparrow\rangle$,
            such that the absolute value of the off-diagonal terms 
            $H^{(4)}_{12}=H^{(4)*}_{21}$ is half of the singlet--triplet
            splitting
            $E^\text{nl}_\text{CT,se}+E^\text{nl}_{J,\text{se}}(\Delta\varphi)$.

\begin{figure}
                \includegraphics{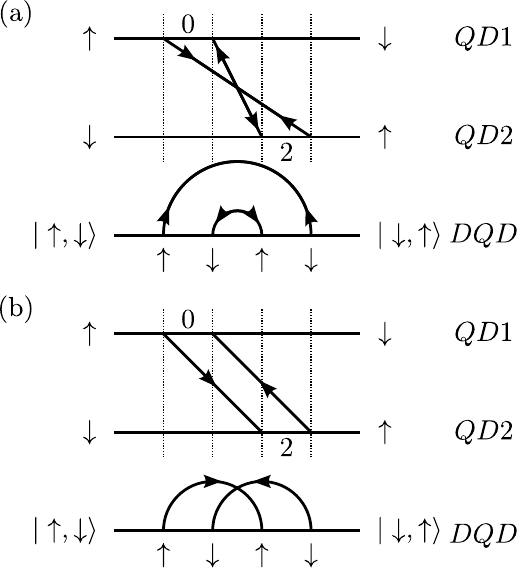}
                \caption{Example diagram of (a) a spin-exchange Josephson and
                    (b) a spin-exchange cotunneling process with the same
                    intermediate QD occupations (denoted by 0 and 2 electrons)
                    read from the left to the right. The upper two horizontal
                    lines represent the QDs and the lower horizontal line
                    represents the DQD as a whole. The spin arrows at the
                    beginning and at the end of the horizontal lines denote the
                    initial state and the final state of the DQD and the arrows
                    on the lead lines indicate the direction of the electrons
                    flowing out of or into the DQD.}
                \label{fig-examplediagram}
            \end{figure}

            To organize all processes, we represent them by diagrams. We take
            the point of view of the DQD system. From this point of view, the
            DQD emits electrons to the leads or absorbs electrons from the
            leads. Due to the excitation gap of the superconductors, tunneling
            proceeds in pairs: the DQD can emit an electron into a
            superconductor which is later reabsorbed, absorb an electron from
            the Fermi sea and subsequently fill the hole which was created,
            emit two electrons which form a Cooper pair, or absorb two
            electrons by destroying a Cooper pair. At this point it does not
            matter which one of the leads enables the process as later on all
            possibilities are summed over.  To keep track of which QD is
            affected by one tunneling event, we represent each QD by one
            horizontal line.  Each tunnel event involving the QD is a vertex
            on this line. A line connecting two vertices indicates, which two
            tunnel events are connected by one of the processes mentioned
            above.  We name it a \emph{lead line}. Two example diagrams are
            shown in Fig.~\ref{fig-examplediagram}. The process on the DQD is
            the same but it is mediated by two different lead processes; in
            Fig.~\ref{fig-examplediagram}(a) the process is mediated by
            Josephson processes whereas it is mediated by cotunneling in
            Fig.~\ref{fig-examplediagram}(b). The direction of the arrows on
            the lead lines indicates the flow of electrons onto or out of the
            QDs.  So the lead lines of Josephson processes have two arrows and
            cotunneling processes have one arrow. The intermediate DQD
            occupations are given by numbers or by small spin arrows.

            Since all processes conserve the total charge of the DQD, they are
            always a sequence of two creation and two anihilation events both
            in the DQD and in the leads. It can easily be checked that all
            possible sequences decompose into a part concerning the leads and a
            part concerning the DQD without acquiring an overall
            fermion-exchange sign. But within both of the subsystems, we need
            to account for possible signs due to fermion exchange. To determine
            the sign of the QD subsystem, the number of permutations is counted
            which would be required to arrange all vertices of QD~1 to the left
            of all vertices of QD~2.  If the number is odd, a fermion-exchange
            sign results.

            \begin{figure*}
                \includegraphics{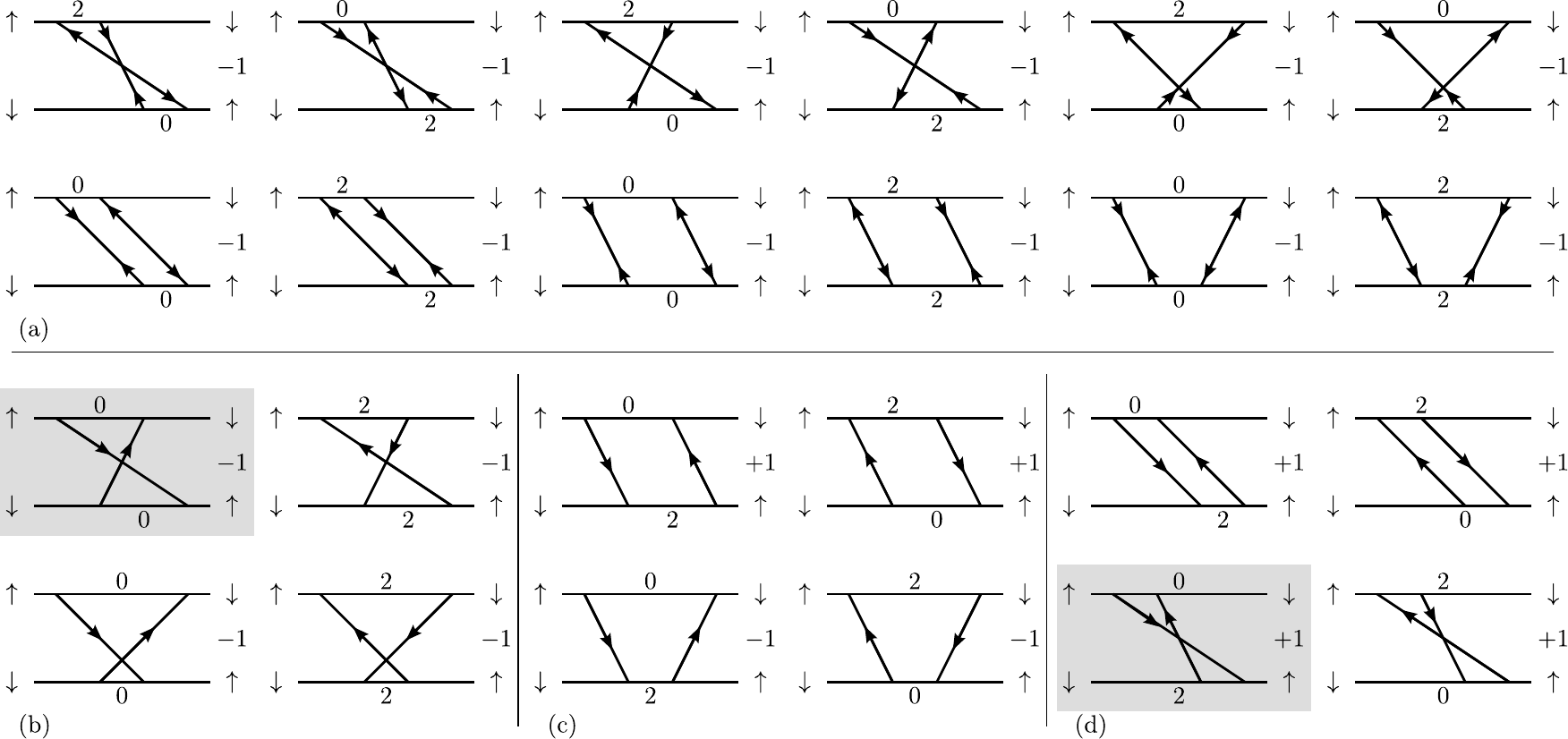}
                \caption{All diagrams contributing to 
                    (a) $\mathcal{M}^{\nonloc}_{J,\text{se}}$ and (b)--(d)
                    $\mathcal{M}^{\nonloc}_\text{CT,se}$.
                    (a) All Josephson processes have a negative overall sign and
                    lower the singlet.
                    (b) Cotunneling processes which lower the singlet. The
                    process of Fig.~\ref{fig-sign}(a) is highlighted.
                    (c) Higher orders of second order cotunneling processes come
                    with different signs but cannot lower the triplet since they
                    are contained in the limit of large $\Delta$.
                    (d) Cotunneling processes which lower the triplet. The
                    process of Fig.~\ref{fig-sign}(b) is highlighted.}
                \label{fig-exchangediagrams}
            \end{figure*}

            If the spin of the electron on a QD is changed in a spin-exchange
            process, another sign may occur. Changing the spin of a QD can be
            done either by removing the electron and filling the QD with an
            electron of opposite spin (intermediate occupation number 0) or by
            adding another electron and removing the first electron afterwards
            (intermediate occupation number 2). In the second case, an
            additional exchange of fermions is necessary when removing the first
            electron, which we call \emph{spin flip via a local singlet}. Such a
            spin flip introduces a sign.

            To determine the sign of the processes on the DQD, we can thus
            summarize the following rules:
            \begin{itemize}
                \item Draw the diagram.
                \item Count the number of permutations which would be required
                    to arrange all vertices of QD~1 to the left of all vertices
                    of QD~2. If it is odd, add a sign.
                \item Count the spin flips via a local singlet (intermediate
                    occupation number 2). Each contributes a fermion-exchange
                    sign.  
            \end{itemize}

            To determine the contributions due to the lead process, we construct
            an auxiliary diagram by collapsing the two lines of the QDs onto
            one.  These auxiliary diagrams are the third horizontal line in
            Figs.~\ref{fig-examplediagram}(a,b). Now each crossing of lead lines
            corresponds to a commutation of lead operators. Furthermore, each
            lead line represents a normal or an anomalous superconducting
            correlation function. If, e.g., a lead line connects two events in
            which, read from the left to the right, first a spin-up electron is
            removed from the superconductors and then a spin-down electron is
            removed from the superconductors, the corresponding correlation 
            function is $\langle c_\downarrow c_\uparrow\rangle$. All
            correlation functions can be calculated using the standard
            Bogoliubov transform \cite{Tinkham1996},
            \begin{subequations}
                \label{eq-leadcorrelations}              
                \begin{align}
                    \langle c_\uparrow^\dagger c_\downarrow^\dagger\rangle&=
                    \langle c_\downarrow c_\uparrow\rangle= 
                    -\langle c_\downarrow^\dagger c_\uparrow^\dagger\rangle=
                    -\langle c_\uparrow c_\downarrow\rangle=
                        \frac{\Delta}{2\sqrt{\varepsilon^2+\Delta^2}}
                    \\
                    \langle c^\dagger_\uparrow c_\uparrow\rangle&=
                    \langle c^\dagger_\downarrow c_\downarrow\rangle=
                    \frac{1}{2}\Big(1-\frac{\varepsilon}
                        {\sqrt{\varepsilon^2+\Delta^2}}\Big)
                    \\
                    \langle c_\uparrow c_\uparrow^\dagger\rangle&=
                    \langle c_\downarrow c_\downarrow^\dagger\rangle=
                    \frac{1}{2}\Big(1+\frac{\varepsilon}
                        {\sqrt{\varepsilon^2+\Delta^2}}\Big),
                \end{align}
            \end{subequations}
            where $\varepsilon$ is the normal-state energy of the lead
            electron measured from the Fermi level. With dispersionful leads
            such as in the wide-band limit, $\varepsilon$ depends on an internal
            quantum number, e.g., $\varepsilon_\mathbf{k}$, which is summed
            over. Note that the order of the spins in the superconducting
            correlation functions is important since the order in which the
            electrons are put into the Cooper pair condensate matters.

            The lead part of the matrix element can thus be obtained by
            following these rules:
            \begin{itemize}
                \item Collapse the two-line diagram to the auxiliary diagram.
                \item Count the number of line crossings. Each crossing
                  contributes a fermion-exchange sign.
                \item Write down the lead correlations following
                  Eq.~\eqref{eq-leadcorrelations}. Take care of fermion-exchange
                  signs that might occur due to the spin-order of Cooper pairs.
                  Use $\varepsilon$ for one lead line and $\varepsilon'$ for the
                  other line.
            \end{itemize}

            Finally, we need to determine the energies of the virtual states.
            The corresponding energies of the DQD can be read off from the
            two-line diagram. The energies of the three virtual states can be
            found by looking at the states in the three spaces between the
            dashed lines in the two line diagrams. Each lead line contributes an
            additional energy $\sqrt{\varepsilon^2+\Delta^2}$ and
            $\sqrt{{\varepsilon'}^2+\Delta^2}$, for the two pairs of tunneling
            events respectively.  
            
            By drawing all diagrams and inserting the corresponding matrix
            elements and the energies into Eq.~\eqref{eq-fourthorderterms}, the
            fourth-order corrections of the ground state energy can be
            constructed explicitly. All of the diagrams of the spin-exchange 
            contributions $\mathcal{M}^\nonloc_{\text{CT}/J,\text{se}}$ are 
            shown in Fig.~\ref{fig-exchangediagrams}. All of them are nonlocal
            since the lead lines connect the two QD lines. From the
            overall exchange signs given next to the diagrams we see that all of
            the Josephson processes [Fig.~\ref{fig-exchangediagrams}(a)] and
            some of the genuine fourth-order cotunneling processes
            [Fig.~\ref{fig-exchangediagrams}(b)] lower the singlet ground state.
            The processes which are second order in electron cotunneling or
            Cooper pair tunneling correspond to diagrams which are reducible in
            the sense that they can be cut into two parts vertically without
            cutting a lead line [Fig.~\ref{fig-exchangediagrams}(c)]. They are
            the processes surviving the limit of large $\Delta$
            (Sec.~\ref{sec-infdelta}) and hence they lower the singlet ground
            state. The remaining cotunneling processes lower the triplet ground
            state [Fig.~\ref{fig-exchangediagrams}(d)].

    \end{appendix}


\begin{thebibliography}{57}%
\makeatletter
\providecommand \@ifxundefined [1]{%
 \@ifx{#1\undefined}
}%
\providecommand \@ifnum [1]{%
 \ifnum #1\expandafter \@firstoftwo
 \else \expandafter \@secondoftwo
 \fi
}%
\providecommand \@ifx [1]{%
 \ifx #1\expandafter \@firstoftwo
 \else \expandafter \@secondoftwo
 \fi
}%
\providecommand \natexlab [1]{#1}%
\providecommand \enquote  [1]{``#1''}%
\providecommand \bibnamefont  [1]{#1}%
\providecommand \bibfnamefont [1]{#1}%
\providecommand \citenamefont [1]{#1}%
\providecommand \href@noop [0]{\@secondoftwo}%
\providecommand \href [0]{\begingroup \@sanitize@url \@href}%
\providecommand \@href[1]{\@@startlink{#1}\@@href}%
\providecommand \@@href[1]{\endgroup#1\@@endlink}%
\providecommand \@sanitize@url [0]{\catcode `\\12\catcode `\$12\catcode
  `\&12\catcode `\#12\catcode `\^12\catcode `\_12\catcode `\%12\relax}%
\providecommand \@@startlink[1]{}%
\providecommand \@@endlink[0]{}%
\providecommand \url  [0]{\begingroup\@sanitize@url \@url }%
\providecommand \@url [1]{\endgroup\@href {#1}{\urlprefix }}%
\providecommand \urlprefix  [0]{URL }%
\providecommand \Eprint [0]{\href }%
\providecommand \doibase [0]{http://dx.doi.org/}%
\providecommand \selectlanguage [0]{\@gobble}%
\providecommand \bibinfo  [0]{\@secondoftwo}%
\providecommand \bibfield  [0]{\@secondoftwo}%
\providecommand \translation [1]{[#1]}%
\providecommand \BibitemOpen [0]{}%
\providecommand \bibitemStop [0]{}%
\providecommand \bibitemNoStop [0]{.\EOS\space}%
\providecommand \EOS [0]{\spacefactor3000\relax}%
\providecommand \BibitemShut  [1]{\csname bibitem#1\endcsname}%
\let\auto@bib@innerbib\@empty
%</preamble>
\bibitem [{\citenamefont {Recher}\ \emph {et~al.}(2001)\citenamefont {Recher},
  \citenamefont {Sukhorukov},\ and\ \citenamefont {Loss}}]{Recher01}%
  \BibitemOpen
  \bibfield  {author} {\bibinfo {author} {\bibfnamefont {P.}~\bibnamefont
  {Recher}}, \bibinfo {author} {\bibfnamefont {E.~V.}\ \bibnamefont
  {Sukhorukov}}, \ and\ \bibinfo {author} {\bibfnamefont {D.}~\bibnamefont
  {Loss}},\ }\href {\doibase 10.1103/PhysRevB.63.165314} {\bibfield  {journal}
  {\bibinfo  {journal} {Phys. Rev. B}\ }\textbf {\bibinfo {volume} {63}},\
  \bibinfo {pages} {165314} (\bibinfo {year} {2001})}\BibitemShut {NoStop}%
\bibitem [{\citenamefont {Lesovik}\ \emph {et~al.}(2001)\citenamefont
  {Lesovik}, \citenamefont {Martin},\ and\ \citenamefont
  {Blatter}}]{Lesovik01}%
  \BibitemOpen
  \bibfield  {author} {\bibinfo {author} {\bibfnamefont {G.}~\bibnamefont
  {Lesovik}}, \bibinfo {author} {\bibfnamefont {T.}~\bibnamefont {Martin}}, \
  and\ \bibinfo {author} {\bibfnamefont {G.}~\bibnamefont {Blatter}},\ }\href
  {\doibase 10.1007/s10051-001-8675-4} {\bibfield  {journal} {\bibinfo
  {journal} {Eur. Phys. J. B}\ }\textbf {\bibinfo {volume} {24}},\ \bibinfo
  {pages} {287} (\bibinfo {year} {2001})}\BibitemShut {NoStop}%
\bibitem [{\citenamefont {Recher}\ and\ \citenamefont {Loss}(2002)}]{Recher02}%
  \BibitemOpen
  \bibfield  {author} {\bibinfo {author} {\bibfnamefont {P.}~\bibnamefont
  {Recher}}\ and\ \bibinfo {author} {\bibfnamefont {D.}~\bibnamefont {Loss}},\
  }\href {\doibase 10.1103/PhysRevB.65.165327} {\bibfield  {journal} {\bibinfo
  {journal} {Phys. Rev. B}\ }\textbf {\bibinfo {volume} {65}},\ \bibinfo
  {pages} {165327} (\bibinfo {year} {2002})}\BibitemShut {NoStop}%
\bibitem [{\citenamefont {Bena}\ \emph {et~al.}(2002)\citenamefont {Bena},
  \citenamefont {Vishveshwara}, \citenamefont {Balents},\ and\ \citenamefont
  {Fisher}}]{Bena02}%
  \BibitemOpen
  \bibfield  {author} {\bibinfo {author} {\bibfnamefont {C.}~\bibnamefont
  {Bena}}, \bibinfo {author} {\bibfnamefont {S.}~\bibnamefont {Vishveshwara}},
  \bibinfo {author} {\bibfnamefont {L.}~\bibnamefont {Balents}}, \ and\
  \bibinfo {author} {\bibfnamefont {M.~P.~A.}\ \bibnamefont {Fisher}},\ }\href
  {\doibase 10.1103/PhysRevLett.89.037901} {\bibfield  {journal} {\bibinfo
  {journal} {Phys. Rev. Lett.}\ }\textbf {\bibinfo {volume} {89}},\ \bibinfo
  {pages} {037901} (\bibinfo {year} {2002})}\BibitemShut {NoStop}%
\bibitem [{\citenamefont {Recher}\ and\ \citenamefont {Loss}(2003)}]{Recher03}%
  \BibitemOpen
  \bibfield  {author} {\bibinfo {author} {\bibfnamefont {P.}~\bibnamefont
  {Recher}}\ and\ \bibinfo {author} {\bibfnamefont {D.}~\bibnamefont {Loss}},\
  }\href {\doibase 10.1103/PhysRevLett.91.267003} {\bibfield  {journal}
  {\bibinfo  {journal} {Phys. Rev. Lett.}\ }\textbf {\bibinfo {volume} {91}},\
  \bibinfo {pages} {267003} (\bibinfo {year} {2003})}\BibitemShut {NoStop}%
\bibitem [{\citenamefont {Prada}\ and\ \citenamefont {Sols}(2004)}]{Prada04}%
  \BibitemOpen
  \bibfield  {author} {\bibinfo {author} {\bibfnamefont {E.}~\bibnamefont
  {Prada}}\ and\ \bibinfo {author} {\bibfnamefont {F.}~\bibnamefont {Sols}},\
  }\href {\doibase 10.1140/epjb/e2004-00284-8} {\bibfield  {journal} {\bibinfo
  {journal} {Eur. Phys. J. B}\ }\textbf {\bibinfo {volume} {40}},\ \bibinfo
  {pages} {379} (\bibinfo {year} {2004})}\BibitemShut {NoStop}%
\bibitem [{\citenamefont {Oliver}\ \emph {et~al.}(2002)\citenamefont {Oliver},
  \citenamefont {Yamaguchi},\ and\ \citenamefont {Yamamoto}}]{Yamaguchi02}%
  \BibitemOpen
  \bibfield  {author} {\bibinfo {author} {\bibfnamefont {W.~D.}\ \bibnamefont
  {Oliver}}, \bibinfo {author} {\bibfnamefont {F.}~\bibnamefont {Yamaguchi}}, \
  and\ \bibinfo {author} {\bibfnamefont {Y.}~\bibnamefont {Yamamoto}},\ }\href
  {\doibase 10.1103/PhysRevLett.88.037901} {\bibfield  {journal} {\bibinfo
  {journal} {Phys. Rev. Lett.}\ }\textbf {\bibinfo {volume} {88}},\ \bibinfo
  {pages} {037901} (\bibinfo {year} {2002})}\BibitemShut {NoStop}%
\bibitem [{\citenamefont {Levy~Yeyati}\ \emph {et~al.}(2007)\citenamefont
  {Levy~Yeyati}, \citenamefont {Bergeret}, \citenamefont {Martin-Rodero},\ and\
  \citenamefont {Klapwijk}}]{Yeyati07}%
  \BibitemOpen
  \bibfield  {author} {\bibinfo {author} {\bibfnamefont {A.}~\bibnamefont
  {Levy~Yeyati}}, \bibinfo {author} {\bibfnamefont {F.~S.}\ \bibnamefont
  {Bergeret}}, \bibinfo {author} {\bibfnamefont {A.}~\bibnamefont
  {Martin-Rodero}}, \ and\ \bibinfo {author} {\bibfnamefont {T.~M.}\
  \bibnamefont {Klapwijk}},\ }\href {\doibase 10.1038/nphys621} {\bibfield
  {journal} {\bibinfo  {journal} {Nat. Phys.}\ }\textbf {\bibinfo {volume}
  {3}},\ \bibinfo {pages} {455} (\bibinfo {year} {2007})}\BibitemShut {NoStop}%
\bibitem [{\citenamefont {Cayssol}(2008)}]{Cayssol08}%
  \BibitemOpen
  \bibfield  {author} {\bibinfo {author} {\bibfnamefont {J.}~\bibnamefont
  {Cayssol}},\ }\href {\doibase 10.1103/PhysRevLett.100.147001} {\bibfield
  {journal} {\bibinfo  {journal} {Phys. Rev. Lett.}\ }\textbf {\bibinfo
  {volume} {100}},\ \bibinfo {pages} {147001} (\bibinfo {year}
  {2008})}\BibitemShut {NoStop}%
\bibitem [{\citenamefont {Sato}\ \emph {et~al.}(2010)\citenamefont {Sato},
  \citenamefont {Loss},\ and\ \citenamefont {Tserkovnyak}}]{Sato10}%
  \BibitemOpen
  \bibfield  {author} {\bibinfo {author} {\bibfnamefont {K.}~\bibnamefont
  {Sato}}, \bibinfo {author} {\bibfnamefont {D.}~\bibnamefont {Loss}}, \ and\
  \bibinfo {author} {\bibfnamefont {Y.}~\bibnamefont {Tserkovnyak}},\ }\href
  {\doibase 10.1103/PhysRevLett.105.226401} {\bibfield  {journal} {\bibinfo
  {journal} {Phys. Rev. Lett.}\ }\textbf {\bibinfo {volume} {105}},\ \bibinfo
  {pages} {226401} (\bibinfo {year} {2010})}\BibitemShut {NoStop}%
\bibitem [{\citenamefont {Schroer}\ \emph {et~al.}(2015)\citenamefont
  {Schroer}, \citenamefont {Silvestrov},\ and\ \citenamefont
  {Recher}}]{Schroer15b}%
  \BibitemOpen
  \bibfield  {author} {\bibinfo {author} {\bibfnamefont {A.}~\bibnamefont
  {Schroer}}, \bibinfo {author} {\bibfnamefont {P.~G.}\ \bibnamefont
  {Silvestrov}}, \ and\ \bibinfo {author} {\bibfnamefont {P.}~\bibnamefont
  {Recher}},\ }\href {\doibase 10.1103/PhysRevB.92.241404} {\bibfield
  {journal} {\bibinfo  {journal} {Phys. Rev. B}\ }\textbf {\bibinfo {volume}
  {92}},\ \bibinfo {pages} {241404} (\bibinfo {year} {2015})}\BibitemShut
  {NoStop}%
\bibitem [{\citenamefont {Hofstetter}\ \emph {et~al.}(2009)\citenamefont
  {Hofstetter}, \citenamefont {Csonka}, \citenamefont {Nyg{\aa}rd},\ and\
  \citenamefont {Sch\"onenberger}}]{Hofstetter09}%
  \BibitemOpen
  \bibfield  {author} {\bibinfo {author} {\bibfnamefont {L.}~\bibnamefont
  {Hofstetter}}, \bibinfo {author} {\bibfnamefont {S.}~\bibnamefont {Csonka}},
  \bibinfo {author} {\bibfnamefont {J.}~\bibnamefont {Nyg{\aa}rd}}, \ and\
  \bibinfo {author} {\bibfnamefont {C.}~\bibnamefont {Sch\"onenberger}},\
  }\href {\doibase 10.1038/nature08432} {\bibfield  {journal} {\bibinfo
  {journal} {Nature (London)}\ }\textbf {\bibinfo {volume} {461}},\ \bibinfo
  {pages} {960} (\bibinfo {year} {2009})}\BibitemShut {NoStop}%
\bibitem [{\citenamefont {Herrmann}\ \emph {et~al.}(2010)\citenamefont
  {Herrmann}, \citenamefont {Portier}, \citenamefont {Roche}, \citenamefont
  {Levy~Yeyati}, \citenamefont {Kontos},\ and\ \citenamefont
  {Strunk}}]{Herrmann10}%
  \BibitemOpen
  \bibfield  {author} {\bibinfo {author} {\bibfnamefont {L.~G.}\ \bibnamefont
  {Herrmann}}, \bibinfo {author} {\bibfnamefont {F.}~\bibnamefont {Portier}},
  \bibinfo {author} {\bibfnamefont {P.}~\bibnamefont {Roche}}, \bibinfo
  {author} {\bibfnamefont {A.}~\bibnamefont {Levy~Yeyati}}, \bibinfo {author}
  {\bibfnamefont {T.}~\bibnamefont {Kontos}}, \ and\ \bibinfo {author}
  {\bibfnamefont {C.}~\bibnamefont {Strunk}},\ }\href {\doibase
  10.1103/PhysRevLett.104.026801} {\bibfield  {journal} {\bibinfo  {journal}
  {Phys. Rev. Lett.}\ }\textbf {\bibinfo {volume} {104}},\ \bibinfo {pages}
  {026801} (\bibinfo {year} {2010})}\BibitemShut {NoStop}%
\bibitem [{\citenamefont {Das}\ \emph {et~al.}(2012)\citenamefont {Das},
  \citenamefont {Ronen}, \citenamefont {Heiblum}, \citenamefont {Mahalu},
  \citenamefont {Kretinin},\ and\ \citenamefont {Shtrikman}}]{Das12}%
  \BibitemOpen
  \bibfield  {author} {\bibinfo {author} {\bibfnamefont {A.}~\bibnamefont
  {Das}}, \bibinfo {author} {\bibfnamefont {Y.}~\bibnamefont {Ronen}}, \bibinfo
  {author} {\bibfnamefont {M.}~\bibnamefont {Heiblum}}, \bibinfo {author}
  {\bibfnamefont {D.}~\bibnamefont {Mahalu}}, \bibinfo {author} {\bibfnamefont
  {A.~V.}\ \bibnamefont {Kretinin}}, \ and\ \bibinfo {author} {\bibfnamefont
  {H.}~\bibnamefont {Shtrikman}},\ }\href {\doibase 10.1038/ncomms2169}
  {\bibfield  {journal} {\bibinfo  {journal} {Nat. Commun.}\ }\textbf {\bibinfo
  {volume} {3}},\ \bibinfo {pages} {1165} (\bibinfo {year} {2012})}\BibitemShut
  {NoStop}%
\bibitem [{\citenamefont {Tan}\ \emph {et~al.}(2015)\citenamefont {Tan},
  \citenamefont {Cox}, \citenamefont {Nieminen}, \citenamefont
  {L\"ahteenm\"aki}, \citenamefont {Golubev}, \citenamefont {Lesovik},\ and\
  \citenamefont {Hakonen}}]{Tan15}%
  \BibitemOpen
  \bibfield  {author} {\bibinfo {author} {\bibfnamefont {Z.~B.}\ \bibnamefont
  {Tan}}, \bibinfo {author} {\bibfnamefont {D.}~\bibnamefont {Cox}}, \bibinfo
  {author} {\bibfnamefont {T.}~\bibnamefont {Nieminen}}, \bibinfo {author}
  {\bibfnamefont {P.}~\bibnamefont {L\"ahteenm\"aki}}, \bibinfo {author}
  {\bibfnamefont {D.}~\bibnamefont {Golubev}}, \bibinfo {author} {\bibfnamefont
  {G.~B.}\ \bibnamefont {Lesovik}}, \ and\ \bibinfo {author} {\bibfnamefont
  {P.~J.}\ \bibnamefont {Hakonen}},\ }\href {\doibase
  10.1103/PhysRevLett.114.096602} {\bibfield  {journal} {\bibinfo  {journal}
  {Phys. Rev. Lett.}\ }\textbf {\bibinfo {volume} {114}},\ \bibinfo {pages}
  {096602} (\bibinfo {year} {2015})}\BibitemShut {NoStop}%
\bibitem [{\citenamefont {F\"ul\"op}\ \emph {et~al.}(2015)\citenamefont
  {F\"ul\"op}, \citenamefont {Dom\'{\i}nguez}, \citenamefont {d'Hollosy},
  \citenamefont {Baumgartner}, \citenamefont {Makk}, \citenamefont {Madsen},
  \citenamefont {Guzenko}, \citenamefont {Nyg\aa{}rd}, \citenamefont
  {Sch\"onenberger}, \citenamefont {Levy~Yeyati},\ and\ \citenamefont
  {Csonka}}]{Fulop15}%
  \BibitemOpen
  \bibfield  {author} {\bibinfo {author} {\bibfnamefont {G.}~\bibnamefont
  {F\"ul\"op}}, \bibinfo {author} {\bibfnamefont {F.}~\bibnamefont
  {Dom\'{\i}nguez}}, \bibinfo {author} {\bibfnamefont {S.}~\bibnamefont
  {d'Hollosy}}, \bibinfo {author} {\bibfnamefont {A.}~\bibnamefont
  {Baumgartner}}, \bibinfo {author} {\bibfnamefont {P.}~\bibnamefont {Makk}},
  \bibinfo {author} {\bibfnamefont {M.~H.}\ \bibnamefont {Madsen}}, \bibinfo
  {author} {\bibfnamefont {V.~A.}\ \bibnamefont {Guzenko}}, \bibinfo {author}
  {\bibfnamefont {J.}~\bibnamefont {Nyg\aa{}rd}}, \bibinfo {author}
  {\bibfnamefont {C.}~\bibnamefont {Sch\"onenberger}}, \bibinfo {author}
  {\bibfnamefont {A.}~\bibnamefont {Levy~Yeyati}}, \ and\ \bibinfo {author}
  {\bibfnamefont {S.}~\bibnamefont {Csonka}},\ }\href {\doibase
  10.1103/PhysRevLett.115.227003} {\bibfield  {journal} {\bibinfo  {journal}
  {Phys. Rev. Lett.}\ }\textbf {\bibinfo {volume} {115}},\ \bibinfo {pages}
  {227003} (\bibinfo {year} {2015})}\BibitemShut {NoStop}%
\bibitem [{\citenamefont {Kawabata}(2001)}]{Kawabata01}%
  \BibitemOpen
  \bibfield  {author} {\bibinfo {author} {\bibfnamefont {S.}~\bibnamefont
  {Kawabata}},\ }\href {\doibase 10.1143/JPSJ.70.1210} {\bibfield  {journal}
  {\bibinfo  {journal} {J. Phys. Soc. Jpn.}\ }\textbf {\bibinfo {volume}
  {70}},\ \bibinfo {pages} {1210} (\bibinfo {year} {2001})}\BibitemShut
  {NoStop}%
\bibitem [{\citenamefont {Chtchelkatchev}\ \emph {et~al.}(2002)\citenamefont
  {Chtchelkatchev}, \citenamefont {Blatter}, \citenamefont {Lesovik},\ and\
  \citenamefont {Martin}}]{Chtchelkatchev02}%
  \BibitemOpen
  \bibfield  {author} {\bibinfo {author} {\bibfnamefont {N.~M.}\ \bibnamefont
  {Chtchelkatchev}}, \bibinfo {author} {\bibfnamefont {G.}~\bibnamefont
  {Blatter}}, \bibinfo {author} {\bibfnamefont {G.~B.}\ \bibnamefont
  {Lesovik}}, \ and\ \bibinfo {author} {\bibfnamefont {T.}~\bibnamefont
  {Martin}},\ }\href {\doibase 10.1103/PhysRevB.66.161320} {\bibfield
  {journal} {\bibinfo  {journal} {Phys. Rev. B}\ }\textbf {\bibinfo {volume}
  {66}},\ \bibinfo {pages} {161320} (\bibinfo {year} {2002})}\BibitemShut
  {NoStop}%
\bibitem [{\citenamefont {Braunecker}\ \emph {et~al.}(2013)\citenamefont
  {Braunecker}, \citenamefont {Burset},\ and\ \citenamefont
  {Levy~Yeyati}}]{Braunecker13}%
  \BibitemOpen
  \bibfield  {author} {\bibinfo {author} {\bibfnamefont {B.}~\bibnamefont
  {Braunecker}}, \bibinfo {author} {\bibfnamefont {P.}~\bibnamefont {Burset}},
  \ and\ \bibinfo {author} {\bibfnamefont {A.}~\bibnamefont {Levy~Yeyati}},\
  }\href {\doibase 10.1103/PhysRevLett.111.136806} {\bibfield  {journal}
  {\bibinfo  {journal} {Phys. Rev. Lett.}\ }\textbf {\bibinfo {volume} {111}},\
  \bibinfo {pages} {136806} (\bibinfo {year} {2013})}\BibitemShut {NoStop}%
\bibitem [{\citenamefont {Burkard}\ \emph {et~al.}(2000)\citenamefont
  {Burkard}, \citenamefont {Loss},\ and\ \citenamefont
  {Sukhorukov}}]{Burkard00}%
  \BibitemOpen
  \bibfield  {author} {\bibinfo {author} {\bibfnamefont {G.}~\bibnamefont
  {Burkard}}, \bibinfo {author} {\bibfnamefont {D.}~\bibnamefont {Loss}}, \
  and\ \bibinfo {author} {\bibfnamefont {E.~V.}\ \bibnamefont {Sukhorukov}},\
  }\href {\doibase 10.1103/PhysRevB.61.R16303} {\bibfield  {journal} {\bibinfo
  {journal} {Phys. Rev. B}\ }\textbf {\bibinfo {volume} {61}},\ \bibinfo
  {pages} {R16303} (\bibinfo {year} {2000})}\BibitemShut {NoStop}%
\bibitem [{\citenamefont {Hu}\ and\ \citenamefont {Das~Sarma}(2004)}]{Hu04}%
  \BibitemOpen
  \bibfield  {author} {\bibinfo {author} {\bibfnamefont {X.}~\bibnamefont
  {Hu}}\ and\ \bibinfo {author} {\bibfnamefont {S.}~\bibnamefont {Das~Sarma}},\
  }\href {\doibase 10.1103/PhysRevB.69.115312} {\bibfield  {journal} {\bibinfo
  {journal} {Phys. Rev. B}\ }\textbf {\bibinfo {volume} {69}},\ \bibinfo
  {pages} {115312} (\bibinfo {year} {2004})}\BibitemShut {NoStop}%
\bibitem [{\citenamefont {Samuelsson}\ \emph {et~al.}(2004)\citenamefont
  {Samuelsson}, \citenamefont {Sukhorukov},\ and\ \citenamefont
  {B\"uttiker}}]{Samuelsson04-2}%
  \BibitemOpen
  \bibfield  {author} {\bibinfo {author} {\bibfnamefont {P.}~\bibnamefont
  {Samuelsson}}, \bibinfo {author} {\bibfnamefont {E.~V.}\ \bibnamefont
  {Sukhorukov}}, \ and\ \bibinfo {author} {\bibfnamefont {M.}~\bibnamefont
  {B\"uttiker}},\ }\href {\doibase 10.1103/PhysRevB.70.115330} {\bibfield
  {journal} {\bibinfo  {journal} {Phys. Rev. B}\ }\textbf {\bibinfo {volume}
  {70}},\ \bibinfo {pages} {115330} (\bibinfo {year} {2004})}\BibitemShut
  {NoStop}%
\bibitem [{\citenamefont {San-Jose}\ and\ \citenamefont
  {Prada}(2006)}]{Prada06}%
  \BibitemOpen
  \bibfield  {author} {\bibinfo {author} {\bibfnamefont {P.}~\bibnamefont
  {San-Jose}}\ and\ \bibinfo {author} {\bibfnamefont {E.}~\bibnamefont
  {Prada}},\ }\href {\doibase 10.1103/PhysRevB.74.045305} {\bibfield  {journal}
  {\bibinfo  {journal} {Phys. Rev. B}\ }\textbf {\bibinfo {volume} {74}},\
  \bibinfo {pages} {045305} (\bibinfo {year} {2006})}\BibitemShut {NoStop}%
\bibitem [{\citenamefont {Mazza}\ \emph {et~al.}(2013)\citenamefont {Mazza},
  \citenamefont {Braunecker}, \citenamefont {Recher},\ and\ \citenamefont
  {Levy~Yeyati}}]{Mazza13}%
  \BibitemOpen
  \bibfield  {author} {\bibinfo {author} {\bibfnamefont {F.}~\bibnamefont
  {Mazza}}, \bibinfo {author} {\bibfnamefont {B.}~\bibnamefont {Braunecker}},
  \bibinfo {author} {\bibfnamefont {P.}~\bibnamefont {Recher}}, \ and\ \bibinfo
  {author} {\bibfnamefont {A.}~\bibnamefont {Levy~Yeyati}},\ }\href {\doibase
  10.1103/PhysRevB.88.195403} {\bibfield  {journal} {\bibinfo  {journal} {Phys.
  Rev. B}\ }\textbf {\bibinfo {volume} {88}},\ \bibinfo {pages} {195403}
  (\bibinfo {year} {2013})}\BibitemShut {NoStop}%
\bibitem [{\citenamefont {Schroer}\ \emph {et~al.}(2014)\citenamefont
  {Schroer}, \citenamefont {Braunecker}, \citenamefont {Levy~Yeyati},\ and\
  \citenamefont {Recher}}]{Schroer14}%
  \BibitemOpen
  \bibfield  {author} {\bibinfo {author} {\bibfnamefont {A.}~\bibnamefont
  {Schroer}}, \bibinfo {author} {\bibfnamefont {B.}~\bibnamefont {Braunecker}},
  \bibinfo {author} {\bibfnamefont {A.}~\bibnamefont {Levy~Yeyati}}, \ and\
  \bibinfo {author} {\bibfnamefont {P.}~\bibnamefont {Recher}},\ }\href
  {\doibase 10.1103/PhysRevLett.113.266401} {\bibfield  {journal} {\bibinfo
  {journal} {Phys. Rev. Lett.}\ }\textbf {\bibinfo {volume} {113}},\ \bibinfo
  {pages} {266401} (\bibinfo {year} {2014})}\BibitemShut {NoStop}%
\bibitem [{\citenamefont {Cerletti}\ \emph {et~al.}(2005)\citenamefont
  {Cerletti}, \citenamefont {Gywat},\ and\ \citenamefont {Loss}}]{Cerletti05}%
  \BibitemOpen
  \bibfield  {author} {\bibinfo {author} {\bibfnamefont {V.}~\bibnamefont
  {Cerletti}}, \bibinfo {author} {\bibfnamefont {O.}~\bibnamefont {Gywat}}, \
  and\ \bibinfo {author} {\bibfnamefont {D.}~\bibnamefont {Loss}},\ }\href
  {\doibase 10.1103/PhysRevB.72.115316} {\bibfield  {journal} {\bibinfo
  {journal} {Phys. Rev. B}\ }\textbf {\bibinfo {volume} {72}},\ \bibinfo
  {pages} {115316} (\bibinfo {year} {2005})}\BibitemShut {NoStop}%
\bibitem [{\citenamefont {Titov}\ \emph {et~al.}(2005)\citenamefont {Titov},
  \citenamefont {Trauzettel}, \citenamefont {Michaelis},\ and\ \citenamefont
  {Beenakker}}]{Titov05}%
  \BibitemOpen
  \bibfield  {author} {\bibinfo {author} {\bibfnamefont {M.}~\bibnamefont
  {Titov}}, \bibinfo {author} {\bibfnamefont {B.}~\bibnamefont {Trauzettel}},
  \bibinfo {author} {\bibfnamefont {B.}~\bibnamefont {Michaelis}}, \ and\
  \bibinfo {author} {\bibfnamefont {C.~W.~J.}\ \bibnamefont {Beenakker}},\
  }\href {http://stacks.iop.org/1367-2630/7/i=1/a=186} {\bibfield  {journal}
  {\bibinfo  {journal} {New J. Phys.}\ }\textbf {\bibinfo {volume} {7}},\
  \bibinfo {pages} {186} (\bibinfo {year} {2005})}\BibitemShut {NoStop}%
\bibitem [{\citenamefont {Budich}\ and\ \citenamefont
  {Trauzettel}(2010)}]{Budich09}%
  \BibitemOpen
  \bibfield  {author} {\bibinfo {author} {\bibfnamefont {J.~C.}\ \bibnamefont
  {Budich}}\ and\ \bibinfo {author} {\bibfnamefont {B.}~\bibnamefont
  {Trauzettel}},\ }\href {\doibase 10.1088/0957-4484/21/27/274001} {\bibfield
  {journal} {\bibinfo  {journal} {Nanotechnology}\ }\textbf {\bibinfo {volume}
  {21}},\ \bibinfo {pages} {274001} (\bibinfo {year} {2010})}\BibitemShut
  {NoStop}%
\bibitem [{\citenamefont {Nigg}\ \emph {et~al.}(2015)\citenamefont {Nigg},
  \citenamefont {Tiwari}, \citenamefont {Walter},\ and\ \citenamefont
  {Schmidt}}]{Nigg14}%
  \BibitemOpen
  \bibfield  {author} {\bibinfo {author} {\bibfnamefont {S.~E.}\ \bibnamefont
  {Nigg}}, \bibinfo {author} {\bibfnamefont {R.~P.}\ \bibnamefont {Tiwari}},
  \bibinfo {author} {\bibfnamefont {S.}~\bibnamefont {Walter}}, \ and\ \bibinfo
  {author} {\bibfnamefont {T.~L.}\ \bibnamefont {Schmidt}},\ }\href {\doibase
  10.1103/PhysRevB.91.094516} {\bibfield  {journal} {\bibinfo  {journal} {Phys.
  Rev. B}\ }\textbf {\bibinfo {volume} {91}},\ \bibinfo {pages} {094516}
  (\bibinfo {year} {2015})}\BibitemShut {NoStop}%
\bibitem [{\citenamefont {Schroer}\ and\ \citenamefont
  {Recher}(2015)}]{Schroer15}%
  \BibitemOpen
  \bibfield  {author} {\bibinfo {author} {\bibfnamefont {A.}~\bibnamefont
  {Schroer}}\ and\ \bibinfo {author} {\bibfnamefont {P.}~\bibnamefont
  {Recher}},\ }\href {\doibase 10.1103/PhysRevB.92.054514} {\bibfield
  {journal} {\bibinfo  {journal} {Phys. Rev. B}\ }\textbf {\bibinfo {volume}
  {92}},\ \bibinfo {pages} {054514} (\bibinfo {year} {2015})}\BibitemShut
  {NoStop}%
\bibitem [{\citenamefont {Deacon}\ \emph {et~al.}(2015)\citenamefont {Deacon},
  \citenamefont {Oiwa}, \citenamefont {Sailer}, \citenamefont {Baba},
  \citenamefont {Kanai}, \citenamefont {Shibata}, \citenamefont {Hirakawa},\
  and\ \citenamefont {Tarucha}}]{Deacon15}%
  \BibitemOpen
  \bibfield  {author} {\bibinfo {author} {\bibfnamefont {R.~S.}\ \bibnamefont
  {Deacon}}, \bibinfo {author} {\bibfnamefont {A.}~\bibnamefont {Oiwa}},
  \bibinfo {author} {\bibfnamefont {J.}~\bibnamefont {Sailer}}, \bibinfo
  {author} {\bibfnamefont {S.}~\bibnamefont {Baba}}, \bibinfo {author}
  {\bibfnamefont {Y.}~\bibnamefont {Kanai}}, \bibinfo {author} {\bibfnamefont
  {K.}~\bibnamefont {Shibata}}, \bibinfo {author} {\bibfnamefont
  {K.}~\bibnamefont {Hirakawa}}, \ and\ \bibinfo {author} {\bibfnamefont
  {S.}~\bibnamefont {Tarucha}},\ }\href {http://dx.doi.org/10.1038/ncomms8446}
  {\bibfield  {journal} {\bibinfo  {journal} {Nat. Commun.}\ }\textbf {\bibinfo
  {volume} {6}},\ \bibinfo {pages} {7446} (\bibinfo {year} {2015})}\BibitemShut
  {NoStop}%
\bibitem [{\citenamefont {Choi}\ \emph {et~al.}(2000)\citenamefont {Choi},
  \citenamefont {Bruder},\ and\ \citenamefont {Loss}}]{Choi00}%
  \BibitemOpen
  \bibfield  {author} {\bibinfo {author} {\bibfnamefont {M.-S.}\ \bibnamefont
  {Choi}}, \bibinfo {author} {\bibfnamefont {C.}~\bibnamefont {Bruder}}, \ and\
  \bibinfo {author} {\bibfnamefont {D.}~\bibnamefont {Loss}},\ }\href {\doibase
  10.1103/PhysRevB.62.13569} {\bibfield  {journal} {\bibinfo  {journal} {Phys.
  Rev. B}\ }\textbf {\bibinfo {volume} {62}},\ \bibinfo {pages} {13569}
  (\bibinfo {year} {2000})}\BibitemShut {NoStop}%
\bibitem [{\citenamefont {Wang}\ and\ \citenamefont {Hu}(2011)}]{Wang11}%
  \BibitemOpen
  \bibfield  {author} {\bibinfo {author} {\bibfnamefont {Z.}~\bibnamefont
  {Wang}}\ and\ \bibinfo {author} {\bibfnamefont {X.}~\bibnamefont {Hu}},\
  }\href {\doibase 10.1103/PhysRevLett.106.037002} {\bibfield  {journal}
  {\bibinfo  {journal} {Phys. Rev. Lett.}\ }\textbf {\bibinfo {volume} {106}},\
  \bibinfo {pages} {037002} (\bibinfo {year} {2011})}\BibitemShut {NoStop}%
\bibitem [{\citenamefont {Pan}\ and\ \citenamefont {Lin}(2006)}]{Pan06}%
  \BibitemOpen
  \bibfield  {author} {\bibinfo {author} {\bibfnamefont {H.}~\bibnamefont
  {Pan}}\ and\ \bibinfo {author} {\bibfnamefont {T.-H.}\ \bibnamefont {Lin}},\
  }\href {\doibase 10.1103/PhysRevB.74.235312} {\bibfield  {journal} {\bibinfo
  {journal} {Phys. Rev. B}\ }\textbf {\bibinfo {volume} {74}},\ \bibinfo
  {pages} {235312} (\bibinfo {year} {2006})}\BibitemShut {NoStop}%
\bibitem [{\citenamefont {Jacquet}\ \emph {et~al.}(2015)\citenamefont
  {Jacquet}, \citenamefont {Rech}, \citenamefont {Jonckheere}, \citenamefont
  {Zazunov},\ and\ \citenamefont {Martin}}]{Jacquet15}%
  \BibitemOpen
  \bibfield  {author} {\bibinfo {author} {\bibfnamefont {R.}~\bibnamefont
  {Jacquet}}, \bibinfo {author} {\bibfnamefont {J.}~\bibnamefont {Rech}},
  \bibinfo {author} {\bibfnamefont {T.}~\bibnamefont {Jonckheere}}, \bibinfo
  {author} {\bibfnamefont {A.}~\bibnamefont {Zazunov}}, \ and\ \bibinfo
  {author} {\bibfnamefont {T.}~\bibnamefont {Martin}},\ }\href {\doibase
  10.1103/PhysRevB.92.235429} {\bibfield  {journal} {\bibinfo  {journal} {Phys.
  Rev. B}\ }\textbf {\bibinfo {volume} {92}},\ \bibinfo {pages} {235429}
  (\bibinfo {year} {2015})}\BibitemShut {NoStop}%
\bibitem [{\citenamefont {Lee}\ \emph {et~al.}(2010)\citenamefont {Lee},
  \citenamefont {Jonckheere},\ and\ \citenamefont {Martin}}]{Lee14}%
  \BibitemOpen
  \bibfield  {author} {\bibinfo {author} {\bibfnamefont {M.}~\bibnamefont
  {Lee}}, \bibinfo {author} {\bibfnamefont {T.}~\bibnamefont {Jonckheere}}, \
  and\ \bibinfo {author} {\bibfnamefont {T.}~\bibnamefont {Martin}},\ }\href
  {\doibase 10.1103/PhysRevB.81.155114} {\bibfield  {journal} {\bibinfo
  {journal} {Phys. Rev. B}\ }\textbf {\bibinfo {volume} {81}},\ \bibinfo
  {pages} {155114} (\bibinfo {year} {2010})}\BibitemShut {NoStop}%
\bibitem [{\citenamefont {Rozhkov}\ and\ \citenamefont
  {Arovas}(1999)}]{Rozhkov1999a}%
  \BibitemOpen
  \bibfield  {author} {\bibinfo {author} {\bibfnamefont {A.~V.}\ \bibnamefont
  {Rozhkov}}\ and\ \bibinfo {author} {\bibfnamefont {D.~P.}\ \bibnamefont
  {Arovas}},\ }\href {\doibase 10.1103/PhysRevLett.82.2788} {\bibfield
  {journal} {\bibinfo  {journal} {Phys. Rev. Lett.}\ }\textbf {\bibinfo
  {volume} {82}},\ \bibinfo {pages} {2788} (\bibinfo {year}
  {1999})}\BibitemShut {NoStop}%
\bibitem [{\citenamefont {Bergeret}\ \emph {et~al.}(2006)\citenamefont
  {Bergeret}, \citenamefont {Yeyati},\ and\ \citenamefont
  {Mart\'{\i}n-Rodero}}]{Bergeret2006a}%
  \BibitemOpen
  \bibfield  {author} {\bibinfo {author} {\bibfnamefont {F.~S.}\ \bibnamefont
  {Bergeret}}, \bibinfo {author} {\bibfnamefont {A.~L.}\ \bibnamefont
  {Yeyati}}, \ and\ \bibinfo {author} {\bibfnamefont {A.}~\bibnamefont
  {Mart\'{\i}n-Rodero}},\ }\href {\doibase 10.1103/PhysRevB.74.132505}
  {\bibfield  {journal} {\bibinfo  {journal} {Phys. Rev. B}\ }\textbf {\bibinfo
  {volume} {74}},\ \bibinfo {pages} {132505} (\bibinfo {year}
  {2006})}\BibitemShut {NoStop}%
\bibitem [{\citenamefont {L\'opez}\ \emph {et~al.}(2007)\citenamefont
  {L\'opez}, \citenamefont {Choi},\ and\ \citenamefont {Aguado}}]{Lopez2007a}%
  \BibitemOpen
  \bibfield  {author} {\bibinfo {author} {\bibfnamefont {R.}~\bibnamefont
  {L\'opez}}, \bibinfo {author} {\bibfnamefont {M.-S.}\ \bibnamefont {Choi}}, \
  and\ \bibinfo {author} {\bibfnamefont {R.}~\bibnamefont {Aguado}},\ }\href
  {\doibase 10.1103/PhysRevB.75.045132} {\bibfield  {journal} {\bibinfo
  {journal} {Phys. Rev. B}\ }\textbf {\bibinfo {volume} {75}},\ \bibinfo
  {pages} {045132} (\bibinfo {year} {2007})}\BibitemShut {NoStop}%
\bibitem [{\citenamefont {Krishna-murthy}\ \emph {et~al.}(1980)\citenamefont
  {Krishna-murthy}, \citenamefont {Wilkins},\ and\ \citenamefont
  {Wilson}}]{Krishna1980a}%
  \BibitemOpen
  \bibfield  {author} {\bibinfo {author} {\bibfnamefont {H.~R.}\ \bibnamefont
  {Krishna-murthy}}, \bibinfo {author} {\bibfnamefont {J.~W.}\ \bibnamefont
  {Wilkins}}, \ and\ \bibinfo {author} {\bibfnamefont {K.~G.}\ \bibnamefont
  {Wilson}},\ }\href {\doibase 10.1103/PhysRevB.21.1003} {\bibfield  {journal}
  {\bibinfo  {journal} {Phys. Rev. B}\ }\textbf {\bibinfo {volume} {21}},\
  \bibinfo {pages} {1003} (\bibinfo {year} {1980})}\BibitemShut {NoStop}%
\bibitem [{\citenamefont {Choi}\ \emph {et~al.}(2004)\citenamefont {Choi},
  \citenamefont {Lee}, \citenamefont {Kang},\ and\ \citenamefont
  {Belzig}}]{Choi2004a}%
  \BibitemOpen
  \bibfield  {author} {\bibinfo {author} {\bibfnamefont {M.-S.}\ \bibnamefont
  {Choi}}, \bibinfo {author} {\bibfnamefont {M.}~\bibnamefont {Lee}}, \bibinfo
  {author} {\bibfnamefont {K.}~\bibnamefont {Kang}}, \ and\ \bibinfo {author}
  {\bibfnamefont {W.}~\bibnamefont {Belzig}},\ }\href {\doibase
  10.1103/PhysRevB.70.020502} {\bibfield  {journal} {\bibinfo  {journal} {Phys.
  Rev. B}\ }\textbf {\bibinfo {volume} {70}},\ \bibinfo {pages} {020502}
  (\bibinfo {year} {2004})}\BibitemShut {NoStop}%
\bibitem [{\citenamefont {Governale}\ \emph {et~al.}(2008)\citenamefont
  {Governale}, \citenamefont {Pala},\ and\ \citenamefont
  {K\"onig}}]{Governale2008a}%
  \BibitemOpen
  \bibfield  {author} {\bibinfo {author} {\bibfnamefont {M.}~\bibnamefont
  {Governale}}, \bibinfo {author} {\bibfnamefont {M.~G.}\ \bibnamefont {Pala}},
  \ and\ \bibinfo {author} {\bibfnamefont {J.}~\bibnamefont {K\"onig}},\ }\href
  {\doibase 10.1103/PhysRevB.77.134513} {\bibfield  {journal} {\bibinfo
  {journal} {Phys. Rev. B}\ }\textbf {\bibinfo {volume} {77}},\ \bibinfo
  {pages} {134513} (\bibinfo {year} {2008})}\BibitemShut {NoStop}%
\bibitem [{\citenamefont {Siano}\ and\ \citenamefont
  {Egger}(2004)}]{Siano2004a}%
  \BibitemOpen
  \bibfield  {author} {\bibinfo {author} {\bibfnamefont {F.}~\bibnamefont
  {Siano}}\ and\ \bibinfo {author} {\bibfnamefont {R.}~\bibnamefont {Egger}},\
  }\href {\doibase 10.1103/PhysRevLett.93.047002} {\bibfield  {journal}
  {\bibinfo  {journal} {Phys. Rev. Lett.}\ }\textbf {\bibinfo {volume} {93}},\
  \bibinfo {pages} {047002} (\bibinfo {year} {2004})}\BibitemShut {NoStop}%
\bibitem [{\citenamefont {Clerk}\ and\ \citenamefont
  {Ambegaokar}(2000)}]{Aashish2000a}%
  \BibitemOpen
  \bibfield  {author} {\bibinfo {author} {\bibfnamefont {A.~A.}\ \bibnamefont
  {Clerk}}\ and\ \bibinfo {author} {\bibfnamefont {V.}~\bibnamefont
  {Ambegaokar}},\ }\href {\doibase 10.1103/PhysRevB.61.9109} {\bibfield
  {journal} {\bibinfo  {journal} {Phys. Rev. B}\ }\textbf {\bibinfo {volume}
  {61}},\ \bibinfo {pages} {9109} (\bibinfo {year} {2000})}\BibitemShut
  {NoStop}%
\bibitem [{\citenamefont {Affleck}\ \emph {et~al.}(2000)\citenamefont
  {Affleck}, \citenamefont {Caux},\ and\ \citenamefont
  {Zagoskin}}]{Affleck2000a}%
  \BibitemOpen
  \bibfield  {author} {\bibinfo {author} {\bibfnamefont {I.}~\bibnamefont
  {Affleck}}, \bibinfo {author} {\bibfnamefont {J.-S.}\ \bibnamefont {Caux}}, \
  and\ \bibinfo {author} {\bibfnamefont {A.~M.}\ \bibnamefont {Zagoskin}},\
  }\href {\doibase 10.1103/PhysRevB.62.1433} {\bibfield  {journal} {\bibinfo
  {journal} {Phys. Rev. B}\ }\textbf {\bibinfo {volume} {62}},\ \bibinfo
  {pages} {1433} (\bibinfo {year} {2000})}\BibitemShut {NoStop}%
\bibitem [{\citenamefont {Vecino}\ \emph {et~al.}(2003)\citenamefont {Vecino},
  \citenamefont {Mart\'{\i}n-Rodero},\ and\ \citenamefont
  {Yeyati}}]{Vecino2003a}%
  \BibitemOpen
  \bibfield  {author} {\bibinfo {author} {\bibfnamefont {E.}~\bibnamefont
  {Vecino}}, \bibinfo {author} {\bibfnamefont {A.}~\bibnamefont
  {Mart\'{\i}n-Rodero}}, \ and\ \bibinfo {author} {\bibfnamefont {A.~L.}\
  \bibnamefont {Yeyati}},\ }\href {\doibase 10.1103/PhysRevB.68.035105}
  {\bibfield  {journal} {\bibinfo  {journal} {Phys. Rev. B}\ }\textbf {\bibinfo
  {volume} {68}},\ \bibinfo {pages} {035105} (\bibinfo {year}
  {2003})}\BibitemShut {NoStop}%
\bibitem [{\citenamefont {Bergeret}\ \emph {et~al.}(2007)\citenamefont
  {Bergeret}, \citenamefont {Yeyati},\ and\ \citenamefont
  {Mart\'{\i}n-Rodero}}]{Bergeret2007a}%
  \BibitemOpen
  \bibfield  {author} {\bibinfo {author} {\bibfnamefont {F.~S.}\ \bibnamefont
  {Bergeret}}, \bibinfo {author} {\bibfnamefont {A.~L.}\ \bibnamefont
  {Yeyati}}, \ and\ \bibinfo {author} {\bibfnamefont {A.}~\bibnamefont
  {Mart\'{\i}n-Rodero}},\ }\href {\doibase 10.1103/PhysRevB.76.174510}
  {\bibfield  {journal} {\bibinfo  {journal} {Phys. Rev. B}\ }\textbf {\bibinfo
  {volume} {76}},\ \bibinfo {pages} {174510} (\bibinfo {year}
  {2007})}\BibitemShut {NoStop}%
\bibitem [{Note1()}]{Note1}%
  \BibitemOpen
  \bibinfo {note} {As the charging state of the QDs must not change, Josephson
  processes and cotunneling processes do not mix to fourth order, cf.~App.~\ref
  {app-pt}}\BibitemShut {NoStop}%
\bibitem [{Note2()}]{Note2}%
  \BibitemOpen
  \bibinfo {note} {An important reason that the critical current differs
  between the singlet phase and the triplet phase is the sign of the nonlocal
  Josephson current. As can be seen from Eqs.~\protect \textup {\hbox
  {\mathsurround \z@ \protect \normalfont (\ignorespaces \ref {eq-epert}\unskip
  \@@italiccorr )}} and~\protect \textup {\hbox {\mathsurround \z@ \protect
  \normalfont (\ignorespaces \ref {eq-lawcos}\unskip \@@italiccorr )}}, it
  depends on the phase, on the parity, and on whether the ground state is a
  singlet or a triplet. So if the local supercurrents and the nonlocal
  supercurrent are flowing in opposite directions at $\Delta \varphi =\pm \pi
  /2$, it can be beneficial to switch to the other ground state at $\Delta
  \varphi '\not =\pm \pi /2$, where the individual supercurrents are smaller
  but flow in the same direction. Due to this interplay it is nontrivial to
  isolate nonlocal features from the critical current.}\BibitemShut {Stop}%
\bibitem [{Note3()}]{Note3}%
  \BibitemOpen
  \bibinfo {note} {Substituting the ground-state energy with the free energy,
  $I=-2ekT/\hbar \times \partial _{\Delta \varphi }\protect \qopname \relax
  o{ln}\DOTSB \sum@ \slimits@ _n e^{-\beta E_n}$, \cite {Bruus04} we verify
  that the total-spin transition is present also at finite temperatures,
  $kT\sim 0.1\Delta $, and hence experimentally accessible.}\BibitemShut
  {Stop}%
\bibitem [{Note4()}]{Note4}%
  \BibitemOpen
  \bibinfo {note} {When integrating out the momentum quantum number in the case
  of continuous leads, at large $\Delta $, second-order Cooper-pair processes
  are independent of $\Delta $ so the relative suppression of genuine
  fourth-order processes may be even stronger.}\BibitemShut {Stop}%
\bibitem [{Note5()}]{Note5}%
  \BibitemOpen
  \bibinfo {note} {It is helpful to consider the equivalent behavior of the
  fully-polarized triplet.}\BibitemShut {Stop}%
\bibitem [{Note6()}]{Note6}%
  \BibitemOpen
  \bibinfo {note} {Here, we neglect the spin-exchange interaction within QD~2.
  Adding it, however, would not change our results because the device is in the
  single-level regime, $\delta \gg U_{2i},t$.}\BibitemShut {Stop}%
\bibitem [{\citenamefont {Recher}\ \emph {et~al.}(2010)\citenamefont {Recher},
  \citenamefont {Nazarov},\ and\ \citenamefont {Kouwenhoven}}]{Recher10}%
  \BibitemOpen
  \bibfield  {author} {\bibinfo {author} {\bibfnamefont {P.}~\bibnamefont
  {Recher}}, \bibinfo {author} {\bibfnamefont {Y.~V.}\ \bibnamefont {Nazarov}},
  \ and\ \bibinfo {author} {\bibfnamefont {L.~P.}\ \bibnamefont
  {Kouwenhoven}},\ }\href {\doibase 10.1103/PhysRevLett.104.156802} {\bibfield
  {journal} {\bibinfo  {journal} {Phys. Rev. Lett.}\ }\textbf {\bibinfo
  {volume} {104}},\ \bibinfo {pages} {156802} (\bibinfo {year}
  {2010})}\BibitemShut {NoStop}%
\bibitem [{\citenamefont {Winkler}(2003)}]{Winkler2003}%
  \BibitemOpen
  \bibfield  {author} {\bibinfo {author} {\bibfnamefont {R.}~\bibnamefont
  {Winkler}},\ }\href@noop {} {\emph {\bibinfo {title} {Spin-orbit coupling
  effects in two-dimensional electron and hole systems}}}\ (\bibinfo
  {publisher} {Springer, Berlin},\ \bibinfo {year} {2003})\BibitemShut
  {NoStop}%
\bibitem [{\citenamefont {Tinkham}(1996)}]{Tinkham1996}%
  \BibitemOpen
  \bibfield  {author} {\bibinfo {author} {\bibfnamefont {M.}~\bibnamefont
  {Tinkham}},\ }\href@noop {} {\emph {\bibinfo {title} {Introduction to
  superconductivity}}},\ \bibinfo {edition} {2nd}\ ed.\ (\bibinfo  {publisher}
  {McGraw-Hill, New York},\ \bibinfo {year} {1996})\BibitemShut {NoStop}%
\bibitem [{\citenamefont {Bruus}\ and\ \citenamefont
  {Flensberg}(2004)}]{Bruus04}%
  \BibitemOpen
  \bibfield  {author} {\bibinfo {author} {\bibfnamefont {H.}~\bibnamefont
  {Bruus}}\ and\ \bibinfo {author} {\bibfnamefont {K.}~\bibnamefont
  {Flensberg}},\ }\href@noop {} {\emph {\bibinfo {title} {Many-Body Quantum
  Theory in Condensed Matter Physics}}}\ (\bibinfo  {publisher} {Oxford
  University Press},\ \bibinfo {year} {2004})\BibitemShut {NoStop}%
\end{thebibliography}
\end{document}